\documentclass[12pt]{iopart}

\usepackage{iopams}  
\usepackage{amsmath}
\usepackage{graphicx}
\usepackage{subfig}
\usepackage{color}
\usepackage{url}
\usepackage{hyperref}

\usepackage{soul}
\usepackage{xcolor}

\RequirePackage[numbers,sort&compress]{natbib}

\begin{document}

\title[Cosmological evolution of collisionless relativistic gases as dark matter]{Cosmological evolution of collisionless kinetic dark matter}

\author{Francisco X. Linares Cede\~no$^{1,a}$, Ulises Nucamendi$^{1,b}$ \& Olivier Sarbach$^{2,c}$}

\address{(1) Facultad de Ciencias F\'isico-Matem\'aticas, Universidad Michoacana de San Nicol\'as de Hidalgo,
Edificio Alfa, Ciudad Universitaria, 58040 Morelia, Michoac\'an, M\'exico.}
\address{(2) Instituto de F\'isica y Matem\'aticas,
Universidad Michoacana de San Nicol\'as de Hidalgo,
Edificio C-3, Ciudad Universitaria, 58040 Morelia, Michoac\'an, M\'exico}
\ead{$^{(a)}$francisco.linares@umich.mx, $^{(b)}$ulises.nucamendi@umich.mx, $^{(c)}$olivier.sarbach@umich.mx}
\vspace{10pt}
\begin{indented}
\item[]\today
\end{indented}

\begin{abstract}
We study a phenomenological dark matter model described as a collisionless relativistic kinetic gas in a spatially flat Friedmann--Lemaître--Robertson--Walker universe. After normalization to the observed present--day dark matter abundance, the model is fully specified by a single dimensionless parameter $\beta$, interpreted as the present particle velocity in units of the speed of light. The resulting energy density, pressure, and sound speed admit closed analytic expressions, interpolating between a radiation--like regime at early times and cold dark matter at late times. We implement the model in a modified version of the Boltzmann code \textsc{CLASS} and confront it with Planck 2018 CMB data. We find that sufficiently small values of $\beta$ are observationally indistinguishable from $\Lambda$CDM, while larger values inducing relativistic effects at early times are constrained. These results establish the consistency of the relativistic kinetic gas scenario with current cosmological observations.
\end{abstract}

%
%
%
%
%

\section{Introduction}

Current cosmological observations are consistent with a pressureless, cold, and collisionless component that dominates the matter budget and drives structure formation on galactic and cosmological scales. This description is strongly supported by measurements of the cosmic microwave background~\cite{Planck:2018vyg}, baryon acoustic oscillations~\cite{BOSS:2016wmc}, and large--scale galaxy clustering~\cite{DES:2021wwk}, which together establish the $\Lambda$CDM model as a remarkably successful framework on linear and mildly non--linear scales.

Nevertheless, increasing observational precision has highlighted a number of open challenges on sub--galactic scales, including discrepancies in the abundance and internal structure of low--mass halos and the diversity of galactic rotation curves~\cite{Weinberg:2013aya}. While these tensions do not invalidate the standard cold dark matter (CDM) paradigm, they motivate the exploration of phenomenological extensions capable of modifying small--scale clustering without spoiling the excellent agreement of $\Lambda$CDM at large scales.

At the same time, the current and forthcoming generation of large--scale structure surveys---such as BOSS, DES, and KiDS~\cite{DES:2021wwk,KiDS:2020suj,BOSS:2016wmc}, as well as upcoming experiments including LSST, Euclid, and the Roman Space Telescope~\cite{LSSTScience:2009jmu,Euclid:2019clj,Eifler:2020vvg}---are delivering high--precision measurements of matter clustering across a wide range of redshifts and scales. These surveys probe not only the two--point statistics of the matter distribution, but increasingly higher--order correlators, redshift--space distortions, and weak--lensing observables, thereby offering novel and stringent tests of the microphysical properties of dark matter~\cite{LSSTDarkMatterGroup:2019mwo,DES:2024lto}.

A particularly well--motivated possibility is that the dark matter component may not be perfectly cold at all times, but instead undergoes a transient relativistic phase during early cosmological evolution. Such behavior arises naturally in scenarios involving warm dark matter~\cite{Bode:2000gq,Viel:2013fqw}, self--interacting dark sectors~\cite{Spergel:1999mh,Tulin:2017ara}, or particles decoupling while semi--relativistic~\cite{Boyanovsky:2007ay,deVega:2009ku}. In these cases, nonzero velocity dispersion and free--streaming effects suppress density perturbations below a characteristic scale, thereby modifying structure formation in a scale--dependent manner. An interesting possibility consists in modeling dark matter through a relativistic kinetic gas. An accurate description of such scenarios requires a consistent relativistic treatment of the dark matter phase--space distribution and its coupling to cosmological perturbations.

The foundations of relativistic continuum dynamics were laid by Synge in his early analysis of the energy--momentum tensor for continuous media~\cite{Synge:1934zzb}. A fully covariant kinetic formulation, based on phase--space distribution functions and applicable to relativistic gases, was later developed in his seminal monograph~\cite{1958PhT....11l..56S}. Building on these ideas, covariant relativistic kinetic theory provides a natural and well--established framework to describe matter components beyond the perfect--fluid approximation~\cite{wI63,CercignaniKremer-Book,Vereshchagin-Book}. In this approach, matter is characterized by a distribution function (DF) defined on the tangent (or cotangent) bundle associated with the spacetime manifold, whose evolution is governed by the general--relativistic Boltzmann (or Vlasov) equation~\cite{1971grc..conf....1E,Stewart-Book,Israel:1976tn,Sarbach:2013uba,rAcGoS2022}. This formalism consistently incorporates relativistic effects, finite velocity dispersion, and departures from local thermodynamic equilibrium, making it particularly well suited for cosmological applications beyond the cold and pressureless limit. For mathematical results regarding the future evolution of cosmological spacetimes involving a Vlasov gas, see for instance Refs.~\cite{Ringstrom-Book,dF16,lAdF20,hBdF22,hA11}.

The dynamics of a collisionless gas is governed by the collisionless Boltzmann (or Vlasov) equation, which ensures conservation of the DF along the geodesic flow generated by the Liouville vector field. Macroscopic quantities such as the particle current and energy--momentum tensor are constructed as suitable momentum--space moments of the DF, allowing for a fully covariant description that incorporates pressure, velocity dispersion, and higher--order kinetic effects in a systematic fashion.

When applied to Friedmann--Lemaître--Robertson--Walker (FLRW) spacetimes, the symmetries associated with homogeneity and isotropy of the background severely restrict the allowed momentum dependence of the DF. Invariance under spatial translations and rotations implies that the most general DF consistent with these symmetries depends only on the cosmic time and on the modulus of the comoving momentum~\cite{Sarbach:2013uba}. Furthermore, for solutions of the Vlasov equation on the future mass
shell, the DF can be written as a function of a conserved quantity $C$, which corresponds to the Euclidean norm of the vector formed by the conserved momentum in each spatial direction. This reduction provides a compact and powerful description of relativistic kinetic matter at the homogeneous and isotropic background level in cosmology.

In the present work, we adopt a particularly simple realization of this framework by modeling the dark matter component as a mono--energetic kinetic gas. Concretely, we consider a single--stream DF sharply peaked at a fixed value of the conserved quantity $C$. Physically, this idealization corresponds to an ensemble of identical particles which, at each time, share the same energy but whose three-momentum is isotropically distributed, giving rise to a nonvanishing kinetic pressure. Note that this approximation is different than the typical thermal DFs (e.g. Maxwell--Boltzmann, Fermi--Dirac, or Bose--Einstein) which arise when collisions dominate. As we will see, it captures the essential ingredients needed to study a smooth transition between relativistic and non--relativistic behavior and allows for fully analytic expressions for the background energy density, pressure, and sound speed.

After normalization to the observed present--day dark matter density, the model is completely characterized by a velocity parameter $\beta$ which, when small enough, can be interpreted as the present particle velocity in units of the speed of light. This parameter controls the redshift at which the transition from radiation--like behavior ($\omega\simeq 1/3$) to CDM--like behavior ($\omega\simeq 0$) occurs, where $\omega$ denotes the ratio between the pressure and the energy density. For small values of $\beta$, the transition takes place at early times and the subsequent cosmological evolution is virtually indistinguishable from standard CDM. Larger values of $\beta$ delay the transition, prolonging the relativistic phase and enhancing free--streaming effects on small scales. In addition, the adiabatic sound speed of the kinetic gas can be computed exactly as a function of $\beta$ and the scale factor, providing direct input for linear cosmological perturbation theory.

To quantify the cosmological viability of this scenario, we incorporate the \textit{Relativistic Kinetic Gas} dark matter (RKG) model into a modified version of the Boltzmann code \textsc{CLASS}~\cite{Lesgourgues:2011re}, allowing us to compute both background evolution and linear perturbations self-consistently. We explore the resulting imprints on the expansion history, the cosmic microwave background (CMB) anisotropies, and the matter power spectrum. Demanding consistency with the observed background cosmology leads to a lower bound on the parameter $\beta$, ensuring that the relativistic phase is effectively confined to sufficiently early epochs. Within the allowed parameter range, the model reproduces the successful large--scale behavior of $\Lambda$CDM while providing a controlled framework to study possible departures from perfect coldness in the dark sector.

This paper is organized as follows. Sec.~\ref{rktg} reviews the covariant kinetic formalism and derives the symmetry--restricted form of the distribution function (DF) in an FLRW spacetime. In Sec.~\ref{sec:monoenergetic}, the mono--energetic RKG model is introduced, and analytic expressions for the energy density, pressure, and equation of state are derived. Constraints on the velocity parameter $\beta$ from Big Bang Nucleosynthesis (BBN) are discussed in Sec.~\ref{sec:Neff_constraint}.  Section~\ref{bg} presents the implementation of the model in \textsc{CLASS} and the resulting background evolution, as well as the corresponding linear perturbations and their impact on cosmological observables. Section~\ref{num_sols} is devoted to the discussion of the numerical solutions obtained from the implementation of the RKG model in the Boltzmann solver \textsc{CLASS}. The statistical analysis and the constraints on the main model parameters from CMB data are presented in Sec.~\ref{sec:statistics}.  Finally, Sec.~\ref{sec:conclusion} contains our conclusions and outlook. Appendix~\ref{app_mass} provides estimates of the dark matter particle mass within the RKG framework. As a complementary analysis, Appendix~\ref{app_lya} presents constraints on the velocity parameter $\beta$ obtained from a dedicated likelihood analysis of Lyman-$\alpha$ forest data.

Throughout this work, we adopt units in which the speed of light is set to unity, $c=1$, unless explicitly stated otherwise.

\section{A brief review of the covariant formulation of the relativistic kinetic theory of gases}
\label{rktg}

We briefly review the basic geometric framework required for the relativistic kinetic theory of gases.  
Let $M$ be an $n$--dimensional spacetime manifold and $T_x M$ the tangent space at a point $x \in M$, whose elements are the tangent vectors $p$ based at $x$. The tangent bundle of $M$ is defined as the disjoint union of all tangent spaces,
\begin{equation}
TM := \bigcup_{x \in M} T_x M \, ,
\label{TM}
\end{equation}
which forms a smooth $2n$--dimensional manifold. The canonical projection $\pi : TM \rightarrow M$ assigns to each $(x,p) \in TM$ its spacetime point $x$. The fiber over a given point $x$, defined as $\pi^{-1}(x)$, is precisely the tangent space at $x$,
\begin{equation}
\pi^{-1}(x) = T_x M \, .
\end{equation}

This structure endows $TM$ with the standard interpretation of a smooth vector bundle over the spacetime manifold $M$.

Local coordinates $x^\mu$ on $M$ induce adapted local coordinates $(x^{\mu}, p^{\mu})$ on the tangent bundle $TM$, where $p^\mu = dx^\mu(p)$. In these coordinates, the natural bases of the tangent and cotangent spaces at a point $(x,p) \in TM$ are respectively given by
\begin{equation}
\left\{ 
    \frac{\partial}{\partial x^{\mu}}, 
    \frac{\partial}{\partial p^{\mu}}
\right\}, 
\qquad
\left\{ 
    dx^{\mu}, 
    dp^{\mu}
\right\}\, ,
\end{equation}
with $\mu = 0,1,\ldots,n$. Thus, any tangent vector $L \in T_{(x,p)}(TM)$ and any cotangent vector 
$\omega \in T^{*}_{(x,p)}(TM)$ admit the expansions
\begin{subequations}
\begin{align}
L &= X^{\mu}\left.\frac{\partial}{\partial x^{\mu}}\right|_{(x,p)}
   + P^{\mu}\left.\frac{\partial}{\partial p^{\mu}}\right|_{(x,p)},
\\[4pt]
\omega &= \alpha_{\mu}\, dx^{\mu}\big|_{(x,p)}
        + \beta_{\mu}\, dp^{\mu}\big|_{(x,p)},
\end{align}
\end{subequations}
where the expansion coefficients are given by
\begin{equation}
X^{\mu} = dx^{\mu}(L), 
\qquad 
P^{\mu} = dp^{\mu}(L),
\qquad
\alpha_{\mu} = \omega\!\left(\frac{\partial}{\partial x^{\mu}}\right),
\qquad
\beta_{\mu} = \omega\!\left(\frac{\partial}{\partial p^{\mu}}\right),
\end{equation}
all evaluated at the phase--space point $(x,p)$.

\subsection{From the Killing algebra of the FLRW spacetime to the collisionless DF}
\label{cdm_df}

We derive the most general DF for a collisionless gas of particles with mass $m$ in a spatially flat FLRW universe. A fully general treatment including non--vanishing spatial curvature can be found in Refs.~\cite{rMsM87a,rMsM87b,rMsM87c,Astorga:2014cka}.

Hence, we consider a spatially flat FLRW spacetime with line element
\begin{equation}
ds^2 = -dt^{2} + a^{2}(t)\left(dx^{2} + dy^{2} + dz^{2}\right)\, ,
\label{FLRW}
\end{equation}
where $a(t)$ is the cosmic scale factor. The metric~\eqref{FLRW} describes a homogeneous and isotropic spacetime whose Killing vector fields are given by~\cite{Astorga:2014cka}
\begin{equation}
\xi(\vec a) = (\vec a \wedge \vec x)\cdot \frac{\partial}{\partial \vec x},
\qquad 
\eta(\vec b) = \vec b \cdot \frac{\partial}{\partial \vec x},
\label{kv}
\end{equation}
for arbitrary constant vectors $\vec a , \vec b \in \mathbb{R}^{3}$, with $\vec x = (x,y,z)$, and where $\wedge$ and $\cdot$ denote the vector and scalar products, respectively. Choosing $\vec a$ and $\vec b$ as the Cartesian basis vectors 
$\{\hat e_1,\hat e_2,\hat e_3\} \equiv \{\hat{\imath},\hat{\jmath},\hat{k}\}$, 
the Killing vector fields~\eqref{kv} take the explicit form~\cite{Astorga:2014cka,Astorga:2017yoj}
\begin{subequations}
\label{kv_basis}
\begin{align}
\xi_1 &= y\frac{\partial}{\partial z} - z\frac{\partial}{\partial y}, 
&\qquad
\xi_2 &= z\frac{\partial}{\partial x} - x\frac{\partial}{\partial z},
&\qquad
\xi_3 &= x\frac{\partial}{\partial y} - y\frac{\partial}{\partial x},
\\[4pt]
\eta_1 &= \frac{\partial}{\partial x},
&\qquad
\eta_2 &= \frac{\partial}{\partial y},
&\qquad
\eta_3 &= \frac{\partial}{\partial z}\, .
\end{align}
\end{subequations}

Here $\xi_i \equiv \xi(\hat e_i)$ and $\eta_i \equiv \eta(\hat e_i)$ for $i=1,2,3$. While the Killing vectors $\xi_i$ are associated with rotational symmetry, in the case of a spatially flat FLRW spacetime the vectors $\eta_i$ generate spatial translations. For non--vanishing spatial curvature, the expressions for $\eta_i$ are more involved, see for instance~\cite{Astorga:2014cka}. The distinction is reflected in the Killing algebra,
\begin{equation}
[\xi_1,\xi_2] = -\xi_3,
\qquad 
[\xi_1,\eta_2] = -\eta_3,
\qquad 
[\eta_1,\eta_2] = 0 \, ,
\end{equation}
(and cyclic permutations) from which $\xi_i$ and $\eta_i$ may be identified as the infinitesimal generators of rotations and translations, respectively, as discussed in Ref.~\cite{Astorga:2014cka}.

We compute the lifted vector fields $\hat{\xi}(\vec a)$ and $\hat{\eta}(\vec b)$ on the tangent bundle $TM$ in adapted local coordinates $(x^{\mu},p^{\mu})$~\cite{Sarbach:2013uba},
\begin{equation}
\hat{\xi}
 = \xi^{\mu}\frac{\partial}{\partial x^{\mu}}
 + p^{\alpha}\frac{\partial \xi^{\mu}}{\partial x^{\alpha}}
   \frac{\partial}{\partial p^{\mu}},
\qquad
\hat{\eta}
 = \eta^{\mu}\frac{\partial}{\partial x^{\mu}}
 + p^{\alpha}\frac{\partial \eta^{\mu}}{\partial x^{\alpha}}
   \frac{\partial}{\partial p^{\mu}} .
\end{equation}
For the Killing vectors given in Eq.~\eqref{kv}, the corresponding lifts read
\begin{equation}
\label{lift_kv}
\hat{\xi}(\vec a)
 = (\vec a \wedge \vec x)\cdot \frac{\partial}{\partial \vec x}
 + (\vec a \wedge \vec p)\cdot \frac{\partial}{\partial \vec p}\, ,
\qquad
\hat{\eta}(\vec b)
 = \vec b \cdot \frac{\partial}{\partial \vec x}\, ,
\end{equation}
where $\vec p=(p^{x},p^{y},p^{z})$. Note that the lifted translation generator $\hat{\eta}$ has the same form as the Killing vector field $\eta$, which is a direct consequence of spatial flatness of the FLRW background.

To determine the DF associated with the metric~\eqref{FLRW}, we consider a smooth function $f: TM \rightarrow \mathbb{R}$ defined on the tangent bundle and invariant under the flows generated by the lifted Killing vector fields $\hat{\xi}(\vec a)$ and $\hat{\eta}(\vec b)$ for all $\vec a,\vec b \in \mathbb{R}^{3}$. Invariance under spatial translations immediately implies that $f$ is independent of the spatial coordinates $\vec x$, such that
\begin{equation}
f = f(t,p^{t},p^{x},p^{y},p^{z}) \, .
\end{equation}

On the other hand, invariance under rotations implies that the DF satisfies
\begin{equation}
(\vec a \wedge \vec p)\cdot \frac{\partial f}{\partial \vec p} = 0
\qquad \text{for all } \vec a \in \mathbb{R}^{3}.
\end{equation}

This condition restricts the momentum dependence of $f$ to rotational invariants, and therefore the DF can be written as  
\begin{equation}
f(x,p) = F\!\big(t,p^{t},|\vec p|\big),
\qquad 
|\vec p| = \sqrt{(p^{x})^{2} + (p^{y})^{2} + (p^{z})^{2}}\, ,
\label{df_FLRW}
\end{equation}
for some smooth function $F$.

At this stage, the form~\eqref{df_FLRW} represents the most general DF compatible solely with the spacetime symmetries of a spatially homogeneous and isotropic FLRW background. No dynamical or microscopic physical assumptions have been imposed so far; the result follows exclusively from the geometric symmetries of the spacetime. In the following, we introduce additional physical requirements on the particle ensemble — such as the specification of the microscopic dynamics and the restriction to the mass shell — which further constrain the functional form of $F$.

For a simple gas of identical particles with positive rest mass $m>0$, it is sufficient to work on the future mass shell $\Gamma^{+}_{m}$, where the temporal momentum component can be eliminated using the mass--shell condition
\begin{equation}
p^{t} = \sqrt{\,m^{2} + a^{2}(t)\,|\vec p|^{2}} \, .
\label{pt_eliminated}
\end{equation}

Now, with the aim of applying this framework to dark matter, we further specialize by assuming that the particles of the gas are collisionless, as is commonly expected for dark matter components.\footnote{This property has been inferred from numerical simulations of structure formation within the cold dark matter paradigm~\cite{Kuhlen:2012ft,Vogelsberger:2019ynw}.
However, the assumption of vanishing pressure—rather than collisionlessness per se—leads to potential tensions at small scales, as it allows the formation of structures down to arbitrarily small scales, in apparent disagreement with current observations~\cite{Weinberg:2013aya}. Models in which dark matter remains collisionless but develops an effective pressure at early times can alleviate these small-scale tensions. In this sense, our RKG model is phenomenologically appealing, as its radiation-like behavior at early times suppresses the formation of small-scale substructures, leading to a cutoff in the matter power spectrum, as will be shown in the following sections.} Under this hypothesis, the DF must satisfy the collisionless Boltzmann (or Vlasov) equation,
\begin{equation}
L[f] 
= p^{\mu}\frac{\partial f}{\partial x^{\mu}}
 - \Gamma^{\mu}_{\;\alpha\beta}
   p^{\alpha}p^{\beta}
   \frac{\partial f}{\partial p^{\mu}}
= 0 \, ,
\label{boltz_eq}
\end{equation}
where $L$ denotes the Liouville operator whose flow describes the lift of the geodesic flow on the tangent bundle.

From Refs.~\cite{Sarbach:2013uba,Astorga:2014cka} it is known that the quantity 
\begin{equation}
Q_{\xi}(x,p) \equiv g_x(p,\xi)
\end{equation}
is conserved along the flow generated by the Liouville vector field $L$, namely,
$L(Q_{\xi})=0$ for any Killing vector field $\xi$. Consequently, the six quantities
\begin{equation}
J_a \equiv Q_{\xi_a} = g(p,\xi_a),
\qquad
K_a \equiv Q_{\eta_a} = g(p,\eta_a),
\qquad a=1,2,3,
\end{equation}
associated with rotational and translational symmetries, respectively, are constants of motion along geodesic trajectories in phase space.

In particular, when there is not spatial curvature (i.e., $k=0$), the rotationally invariant combination is given by
\begin{equation}
C^{2} \equiv |\vec K|^{2} = \sum_{a=1}^{3} K_{a}^{2}\, ,
\end{equation}
which for the spatially flat FLRW metric takes the simple form
\begin{equation}
C(t,|\vec p|) = a^2(t)\,|\vec p| \, .
\end{equation}
Using Eqs.~\eqref{df_FLRW} and~\eqref{pt_eliminated}, together with the conservation of $C$, it follows that the DF may be written as
\begin{equation}
f(x,p) = F(t,m,C),
\label{df_tmC}
\end{equation}
where $C$ is the only nontrivial conserved momentum–space variable compatible with the spacetime symmetries. Physically, the combination $a(t)^2 |\vec p|$ represents the (Euclidean) norm of the vector formed by the conserved momentum components $K_i = p_i = a^2(t) p^i$ in each spatial direction $\hat{e}_i$. Note that it differs from the magnitude $a(t)|\vec p|$ of the physical 3--momentum of a particle as measured by comoving observers by a factor $a(t)$.

Substituting the expression~\eqref{df_tmC} into the Boltzmann equation~\eqref{boltz_eq}, one finds that the Liouville operator reduces to
\begin{equation}
p^{t}\,\frac{\partial f}{\partial t} = 0 \, ,
\end{equation}
from which it follows that the DF is independent of time and can be written as
\begin{equation}
f(x,p) = F(m,C) \, .
\label{dist_funct}
\end{equation}

Therefore, for a collisionless gas of cold dark matter particles evolving in a spatially flat FLRW spacetime, the most general DF compatible with both the spacetime symmetries and the collisionless dynamics depends only on the particle mass $m$ and on the conserved quantity $C = a^2(t)|\vec p|$.

\subsection{Physical quantities of the Relativistic Kinetic Gas}
\label{physical_quantities}

The particle number current density $J^{\mu}$ and the energy--momentum tensor $T^{\mu\nu}$ associated with a collisionless relativistic gas with distribution function $f(x,p)$ are defined covariantly as~\cite{Sarbach:2013uba,Sarbach:2013fya}
\begin{equation}
J^{\mu}(x) = \int_{P_x^+(m)} f(x,p)\, p^{\mu} \, \mbox{dvol}_x(p), \qquad T^{\mu\nu}(x) = \int_{P_x^+(m)} f(x,p)\, p^{\mu}p^{\nu} \, \mbox{dvol}_x(p) \, ,
\end{equation}
where $P_x^+(m)$ denotes the future mass hyperboloid and $ \mbox{dvol}_x(p) = \sqrt{-\det(g_{\mu\nu})} d^3p/p^t $ is the Lorentz-invariant volume element on it.

Specializing these expressions to a spatially flat FLRW geometry and to the DF $ f(x,p)=F\!\big(a^2(t)|\vec p|\big) $ and decomposing $\vec{p} = |\vec p|(\cos\phi\sin\theta,\sin\phi\sin\theta,\cos\theta)$ in terms of spherical coordinates, the above definitions take the explicit form
\begin{align}
J^{\mu}(t) &= \int_{0}^{\infty}\!\!\int_{0}^{\pi}\!\!\int_{0}^{2\pi}
F\!\big(a^2(t)|\vec p|\big)\, p^{\mu}\,
a^3(t)\frac{|\vec p|^{2}\sin\theta \, d\phi\, d\theta\, d|\vec p|}
{\sqrt{m^{2}+a^2(t)|\vec p|^{2}}}\ , , \label{J}
\\[6pt]
T^{\mu\nu}(t) &= \int_{0}^{\infty}\!\!\int_{0}^{\pi}\!\!\int_{0}^{2\pi}
F\!\big(a^2(t)|\vec p|\big)\, p^{\mu}p^{\nu}\,
a^3(t)\frac{|\vec p|^{2}\sin\theta \, d\phi\, d\theta\, d|\vec p|}
{\sqrt{m^{2}+a^2(t)|\vec p|^{2}}} \, .\label{Tmunu}
\end{align}

The physically relevant components of Eqs.~\eqref{J} and~\eqref{Tmunu} take the form
\begin{align}
n(t) &\equiv -J^{\mu} u_{\mu} = \frac{4\pi}{a^{3}(t)} \int_{0}^{\infty} F(C)\, C^{2}\, dC \, , 
\\[6pt]
\rho(t) &\equiv u_\mu u_\mu T^{\mu\nu} = \frac{4\pi}{a^{4}(t)}\int_{0}^{\infty} F(C)\,\sqrt{m^{2}a^{2}(t) + C^{2}} \, C^{2}\, dC \, , \label{rho_gen_1}
\\[6pt]
P(t) &\equiv \frac{1}{3}\, (g_{\mu\nu} + u_\mu u_\nu)T^{\mu\nu} = \frac{4\pi}{3a^{4}(t)} \int_{0}^{\infty} F(C)\, \frac{C^{4}}{\sqrt{m^{2}a^{2}(t) + C^{2}}}\, dC \, ,\label{p_gen_1}
\end{align}
where $n(t)$ is the particle number density, $\rho(t)$ is the energy density, $P(t)$ is the isotropic pressure, and $u^{\mu}=(1,0,0,0)$ denotes the four--velocity of comoving observers.

Notice that Eqs.~\eqref{rho_gen_1} and~\eqref{p_gen_1} can be combined to obtain the exact relation
\begin{equation}
P(t) = \frac{\rho(t)}{3} - \frac{4\pi m^{2}}{3a^{2}(t)}\int_{0}^{\infty} F(C) \frac{C^{2}}{\sqrt{m^{2}a^{2}(t) + C^{2}}}dC\, .
\label{Prho_relation}
\end{equation}

This motivates writing the pressure as
\begin{equation}
P(t) = \frac{\rho(t)}{3} - \Delta P(t), \qquad \Delta P(t) \equiv \frac{4\pi m^{2}}{3a^{2}(t)}\int_{0}^{\infty} F(C)\, \frac{C^{2}}{\sqrt{m^{2}a^{2}(t)+C^{2}}}\, dC \, ,
\end{equation}
where the first term, $\rho/3$, corresponds to the pressure of an {\it ultra--relativistic gas}, i.e.\ radiation--like matter.

The second contribution, $\Delta P$, is strictly positive for any physical
DF $F(C)\ge 0$ and therefore represents a correction that reduces the pressure below the ultra--relativistic value. As a direct consequence, the pressure satisfies the inequality
\begin{equation}
0 \le P \le \frac{\rho}{3}\, .
\label{Eq:PrhoBound}
\end{equation}

The upper bound $P=\rho/3$ is attained in the ultra--relativistic regime, where particle momenta dominate over the rest mass. The lower bound $P\to 0$ is approached in the strictly non--relativistic limit, provided the distribution $F(C)$ is regular and sufficiently localized at small values of $C$.

Since the DF is non--negative, the energy density and pressure obey $\rho \ge 0$ and $P \ge 0$, which implies the validity of the null, weak, and strong energy conditions. Furthermore, as a result of the bound~(\ref{Eq:PrhoBound}), the dominant energy condition,
\begin{equation}
\rho \ge |P|\, ,
\end{equation}
holds as well. These properties are expected on general grounds from the theory of relativistic kinetic theory (see~\cite{1971grc..conf....1E,Sarbach:2013fya} and references therein).

\subsection{Relativistic and non--relativistic limits}

The physical behavior encoded in the relation between $P$ and $\rho$ can be clarified by considering the asymptotic limits of the integrals entering Eqs.~\eqref{rho_gen_1} and~\eqref{p_gen_1}.

\paragraph*{Ultra--relativistic regime.}
When the dominant contribution to the integrals comes from momenta satisfying $C \gg m a(t)$, the mass term can be neglected and $\sqrt{m^{2}a^{2}+C^{2}}\simeq C$. In this limit the energy density and the pressure reduce to
\begin{equation}
\rho(t) \simeq \frac{4\pi}{a^{4}(t)}\int_{0}^{\infty} F(C)\,C^{3} dC,
\qquad P(t) \simeq \frac{4\pi}{3a^{4}(t)}\int_{0}^{\infty} F(C)\,C^{3} dC\, .
\end{equation}

Therefore,
\begin{equation}
P(t) \simeq \frac{\rho(t)}{3}\, ,
\end{equation}
which corresponds to the equation of state of radiation. In this regime the energy density scales as $\rho \propto a^{-4}$, as expected for a relativistic species freely streaming in an expanding universe.

The massless case $m=0$ is naturally included in this limit: since the correction term $\Delta P$ is proportional to $m^{2}$, it vanishes identically when $m=0$, and the exact relation $P=\rho/3$ is recovered, as appropriate for a purely radiation--like fluid.

\paragraph*{Non--relativistic regime.}

When the support of $F$ is dominated by momenta satisfying $C\ll m a(t)$, we may expand the square root as
\begin{equation}
\sqrt{m^{2}a^{2}+C^{2}} \simeq m a \left(1 + \frac{C^{2}}{2 m^{2} a^{2}} + \cdots \right)\, .
\end{equation}

Substituting this approximation into Eqs.~\eqref{rho_gen_1} and~\eqref{p_gen_1}, the leading contributions become
\begin{align}
\rho(t) &\simeq \frac{4\pi m}{a^{3}(t)}\int_{0}^{\infty} F(C)\, C^{2} dC\, ,
\\[4pt]
P(t) &\simeq \frac{4\pi}{3 m\, a^{5}(t)}\int_{0}^{\infty} F(C)\, C^{4} dC \, .
\end{align}

Thus, in the nonrelativistic limit the energy density scales as $\rho \propto a^{-3}$, as appropriate for pressureless matter, whereas the pressure decays faster, $P \propto a^{-5}$, becoming dynamically negligible at late times.

\subsection{Moments of the DF and scale--factor dependence}

It is convenient to express the macroscopic quantities $\rho(t)$ and $P(t)$ in terms of moments of the DF $F(C)$, defined as
\begin{equation}
M_{n} \equiv \int_{0}^{\infty} F(C)\, C^{n}\, dC \, ,
\end{equation}
which are assumed to be finite for the physically relevant values of $n$.

In terms of these moments, the leading contributions to the energy density and pressure in the relativistic and non--relativistic regimes take particularly simple forms. In the ultra--relativistic limit $C \gg m a(t)$, one finds
\begin{equation}
\rho(t) \simeq \frac{4\pi}{a^{4}(t)} M_{3}, \qquad P(t) \simeq \frac{4\pi}{3a^{4}(t)} M_{3}\, .
\end{equation}

Thus, both quantities scale as $a^{-4}$ and satisfy $P\simeq \rho/3$, independently of the detailed shape of $F(C)$.

In the non--relativistic regime $C \ll m a(t)$, the dominant contributions become
\begin{equation}
\rho(t) \simeq \frac{4\pi m}{a^{3}(t)} M_{2}, \qquad P(t) \simeq \frac{4\pi}{3m\,a^{5}(t)} M_{4}\, ,
\end{equation}
so that the energy density exhibits the matter--like scaling $\rho\propto a^{-3}$, while the pressure decays as $P\propto a^{-5}$ and rapidly becomes negligible.

The transition between the relativistic and non--relativistic regimes is governed by the relative magnitude of the rest--mass scale $m a(t)$ compared to the typical momentum scale encoded in the moments of the DF. If $C_{*}$ denotes the characteristic comoving momentum of $F$, then the transition occurs approximately when $m a(t) \sim C_{*}$. This defines a characteristic redshift
\begin{equation}
z_{\rm tr} \sim \frac{m}{C_{*}} - 1\, ,
\label{z_tr}
\end{equation}
marking the epoch at which the kinetic energy of the particles changes from relativistic to nonrelativistic dominance.

Therefore, the detailed form of $F(C)$ determines not only the overall amplitudes of $\rho$ and $P$, through the numerical values of the moments $M_n$, but also the timing of the relativistic--to--nonrelativistic transition and hence the cosmological impact of the dark matter population.

\section{Cosmological implications and mono--energetic relativistic gas}
\label{sec:monoenergetic}

The equation of state resulting from the kinetic description presented above has direct consequences for the cosmological evolution of the dark matter component. Since the pressure always satisfies $0 \le P \le \rho/3$, the kinetic gas interpolates dynamically between a radiation--like behavior at early times and an effectively dust--like behavior at late times.

As we mentioned above, there is an epoch of transition determined by Eq.~\eqref{z_tr} highlighting how the microscopic features of the dark matter distribution determine its macroscopic cosmological evolution. During and prior to this transition, the non--negligible pressure and velocity dispersion lead to free--streaming effects which suppress the growth of matter perturbations below a characteristic comoving scale. 
The magnitude of this suppression depends on the detailed shape of $F(C)$, particularly on its higher moments controlling the pressure and sound speed of the fluid. Broad momentum distributions or relatively large characteristic momenta delay the approach to the non--relativistic regime, increasing the free--streaming length and enhancing the suppression of power on small scales.

In contrast, distributions sharply peaked at low $C$ yield an early transition, rendering the fluid effectively cold at high redshift and minimizing deviations from the standard cold dark matter scenario. Thus, the kinetic description developed here provides a unified framework encompassing both warm-- and cold--dark--matter--like behaviors, with the specific phenomenology dictated by the parameters and shape of the underlying DF.

When the DF is chosen to be mono--energetic,
\begin{equation}
F(C) = \frac{N}{4\pi C^{2}}\,\delta(C-C_{0})\, ,
\label{df_delta}
\end{equation}
the quantities $n$, $\rho$, and $P$ measured by isotropic observers are given by
\begin{equation}\label{n_rho_p}
n(t) = \frac{N}{a^{3}(t)}, \qquad \rho(t) = \frac{N}{a^{4}(t)}\sqrt{m^{2}a^{2}(t)+C_{0}^{2}}, \qquad P(t) = \frac{N}{3a^{4}(t)}\,
\frac{C_{0}^{2}}{\sqrt{m^{2}a^{2}(t)+C_{0}^{2}}}\, .
\end{equation}

The normalization constant $N$ in~\eqref{df_delta} is dimensionless and represents the conserved number of particles per unit comoving volume.  This particular choice of Eq.~\eqref{df_delta} is motivated by a simplified physical picture in which the dark matter particles populate a single--stream phase--space configuration, characterized by a unique comoving momentum scale $C_{0}$. The distribution is sharply localized in the magnitude of the momentum, while remaining isotropic in momentum direction. Since the dynamics is governed by the collisionless Vlasov equation, there is no requirement for the distribution function to be thermal, and non--thermally broadened configurations of this type arise naturally in cosmological settings. In this approximation, all particles at a given spacetime point share the same velocity magnitude, so that the phase--space distribution is sharply localized rather than thermally broadened.

Such a description is appropriate for exploring the minimal effects of velocity dispersion and free--streaming, allowing one to parametrically control the departure from the cold--dark--matter limit through the single parameter $C_{0}$. Larger values of $C_{0}$ correspond to higher characteristic velocities, leading to extended free--streaming lengths and enhanced suppression of structure on small scales, while the limit $C_{0}\to 0$ smoothly reproduces the pressureless cold--dark--matter behavior.

This mono--energetic model does not assume thermal equilibrium and therefore differs conceptually from more conventional choices of DFs, such as Maxwell--Boltzmann, Fermi--Dirac or Bose--Einstein spectra, which describe equilibrated gases with finite temperature and broad momentum distributions. Instead, The delta--function distribution provides an effective description of a collisionless ensemble whose velocity dispersion is dynamically set rather than thermally determined, and serves as a useful toy model for isolating the impact of kinematic effects on cosmological evolution without introducing additional thermal parameters. Since the dynamics is governed by the  Vlasov equation, no thermal equilibration is expected, and non--thermal distribution functions of this type arise naturally.

\subsection{Physical interpretation of $C_0$ in the context of Dark Matter}\label{phys_interp}

Recall that by definition the conserved quantity $C$ is equal to the scale factor times the magnitude $|\vec p_{\rm phys}|$ of the physical $3$-momentum measured by comoving observers. Since in our model all particles have the same value $C_0$ of $C$, this implies that $|\vec p_{\rm phys}|$ decreases in time as $C_0/a(t)$ as the scale factor grows, conformal to the expectation that the gas cools down. The relation between physical momentum and the 3--velocity $\vec v$ measured by comoving observers is the fully relativistic identity
\begin{equation}
\vec p_{\rm phys} = m\,\gamma\, \vec v , \qquad \gamma \equiv \frac{1}{\sqrt{1-|\vec v|^{2}}}\, .
\end{equation}

Introducing the particle's speed $v_0 \equiv \left| \vec v \right|_{a=a_0}$ at the present time and normalizing $a_0=1$, one finds
\begin{equation}
C_0 = m\gamma_0 v_0,\qquad
\gamma_0 \equiv \frac{1}{\sqrt{1-v_0^2}}\, .
\end{equation}

For convenience and physical transparency we introduce the effective velocity parameter
\begin{equation}
\beta \equiv \gamma_0 v_0 \, ,
\label{beta_beta0}
\end{equation}
which allows us to write compactly
\begin{equation}
C_0 = m\, \beta\, .
\end{equation}

Substituting $C_0 = m\, \beta$ into the expressions~\eqref{n_rho_p} for the energy density and pressure yields
\begin{equation}
\rho(a) = \frac{N m }{a^{4}}\sqrt{\,a^{2} + \beta^{2}}, \qquad P(a) = \frac{\rho(a)}{3} \frac{\beta^{2}}{a^{2}+\beta^{2}} \, .
\label{rhoP_beta}
\end{equation}

The parameter $\beta$ provides a unified relativistic characterization of the microscopic dark--matter velocities:

\begin{itemize}
\item In the ultra--relativistic limit ($v_0 \to 1$), one has $\beta\to\infty$ and therefore $P=\rho/3,$ recovering the radiation equation of state.

\item In the non--relativistic regime ($v_0\ll1$), $\beta\simeq\beta_0$, so that at late times ($a\gg\beta$) the pressure decays as
\begin{equation}
P \simeq \frac{\rho}{3}\frac{\beta^{2}}{a^{2}}\propto a^{-5}\, ,
\end{equation}
consistently approaching the dust limit $P\to 0$.
\end{itemize}

From Eq.~\eqref{rhoP_beta} the effective equation--of--state parameter is
\begin{equation}
w(a) \equiv \frac{P}{\rho} = \frac{1}{3}\,\frac{\beta^{2}}{a^{2}+\beta^{2}}\, ,
\label{eos_dm}
\end{equation}
which smoothly interpolates between radiation behavior at early times and cold matter at late times.  The entire cosmological evolution of the dark--matter fluid is therefore controlled by the single parameter $\beta$.

According to Eq.~\eqref{rhoP_beta}, the background evolution of the dark--matter component in the present model is fully characterized by three quantities: the particle mass $m$, the normalization constant $N$ of the mono--energetic distribution, and the effective velocity parameter $\beta$, which encodes the microscopic kinetic properties of the flow.

Considering this kinetic gas as the dark matter componet of the universe, the normalization constant $N$ can be fixed by matching the present--day dark matter energy density,
\begin{equation}
\rho_{DM}(a_0=1) \equiv \rho_{0,DM}\, ,
\end{equation}
so that, using Eq.~\eqref{rhoP_beta}, one obtains
\begin{equation}
\rho_{0,DM} = N m \sqrt{1+\beta^{2}} \qquad \Rightarrow \qquad N = \frac{\rho_{0,DM}}{m\sqrt{1+\beta^{2}}}\, .
\label{m_rho}
\end{equation}

Substituting this expression for $N$ back into Eq.~\eqref{rhoP_beta} yields
the energy density at arbitrary scale factor,
\begin{equation}
\label{rho_dm}
\rho_{DM}(a) = \frac{\rho_{0,DM}}{a^{4}}\sqrt{\frac{a^{2}+\beta^{2}}{1+\beta^{2}}} \, .
\end{equation}

Therefore, once the present value $\rho_{0,DM}$ is fixed by cosmological
observations, the dark--matter background evolution depends on a
\emph{single additional parameter}, namely the effective velocity
$\beta$.

Our aim is to explore the range of values of the effective velocity parameter $\beta$ for which the mono--energetic particle model provides a cosmologically consistent description of the dark--matter component in which the particles behave as relativistic at early times and transition to a cold, pressureless regime at late epochs.

\section{Big Bang Nucleosynthesis constraints from the Effective Number of Relativistic Species}
\label{sec:Neff_constraint}

In the early Universe, any component that behaves as a relativistic fluid contributes to the total radiation density and is therefore constrained by bounds on the effective number of relativistic species, \(N_{\rm eff}\), derived from Big Bang Nucleosynthesis (BBN) and the Cosmic Microwave Background (CMB).

In the early-time limit \(a \ll \beta\), the fluid behaves as radiation (see Eq.~\eqref{eos_dm}). Then,
\begin{equation}
w(a) \longrightarrow \frac{1}{3}\, ,
\end{equation}
and the energy density~\eqref{rho_dm} simplifies to
\begin{equation}
\rho_{DM}(a) \;\xrightarrow{a\ll\beta}\; \Omega_{0,DM}\,\rho_{0,c}\,a^{-4}
\frac{\beta}{\sqrt{1+\beta^2}}\, ,
\end{equation}
where $\rho_{0,c}=3H^2/8\pi G$ is the critical energy density, and $\Omega_{0,DM}\equiv \rho_{0,DM}/\rho_{0,c}$. Thus, the RKG contributes an additional relativistic component during the radiation-dominated era.

The contribution of the RKG model to the effective number of relativistic species is defined as~\cite{Mangano:2005cc,Lesgourgues:2012uu}
\begin{equation}
\Delta N_{\rm eff} = \frac{8}{7}\left(\frac{11}{4}\right)^{4/3}\frac{3P_{\rm RKG}}{\rho_\gamma},
\end{equation}
where $\rho_\gamma = \Omega_{0,\gamma} \rho_{0,c} a^{-4}$ is the photon energy density. In the relativistic regime, where $3P_{\rm RKG} = \rho_{\rm RKG}$, this expression reduces to
\begin{equation}
\Delta N_{\rm eff} = \frac{8}{7}\left(\frac{11}{4}\right)^{4/3}\frac{\rho_{\rm RKG}}{\rho_\gamma}\, ,
\end{equation}
and after substituting the early-time energy density of the dark matter component given by our RKG model yields
\begin{equation}
\Delta N_{\rm eff} = \frac{8}{7}\left(\frac{11}{4}\right)^{4/3}\frac{\Omega_{0,DM}}{\Omega_{0,\gamma}}\frac{\beta}{\sqrt{1+\beta^2}}\, .
\label{eq:Neff_beta_exact}
\end{equation}

Using the observational bound on the total effective number of relativistic degrees of freedom $N_{\rm eff}^{\rm tot} < 3.15$ at $95\%$ C.L.~\cite{Pitrou:2018cgg,Moreno:2025wki}, we separate the total contribution into the standard neutrino background, which contribute with $N_{eff}^{\nu} = 3.046$~\cite{Planck:2018vyg}, and our additional relativistic specie given by the kinetic gas. Thus the RKG model satisfies the upper bound
\begin{equation}
\Delta N_{\rm eff} < 0.104\, ,
\end{equation}
which together with the present-day values $\Omega_{0,DM} \simeq 0.26$ and $\Omega_{0,\gamma} \simeq 5.38\times 10^{-5}$~\cite{Planck:2018vyg}, leads to the upper limit
\begin{equation}
\beta \;\lesssim\; 4.89\times 10^{-6},
\end{equation}
or equivalently,
\begin{equation}
\log_{10}\beta \;\lesssim\; -5.31\, .
\label{logbeta_bbn}
\end{equation}

This constraint arises entirely from the relativistic behavior of the RKG at early times and ensures consistency with BBN and CMB bounds. Notably, the contribution to $\Delta N_{\rm eff}$ is transient and automatically suppressed once $a \gg \beta$, distinguishing this scenario from models with persistent dark radiation. Therefore, the stringent BBN bound on $\Delta N_{\rm eff}$ implies that the parameter $\beta$ must be sufficiently small such that the relativistic regime of the RKG is confined to very early times. With this result, we can safely consider from Eq.~\eqref{beta_beta0} that $\beta \simeq v_0\, .$

\section{Background and linear perturbations evolution}
\label{bg}

The background dynamics of a spatially flat FLRW universe are governed by the Friedmann equations
\begin{eqnarray}
H^{2} &=& \frac{\kappa^{2}}{3}\left(\frac{\rho_{0,b}}{a^{3}} + \frac{\rho_{0,r}}{a^{4}} + \frac{\rho_{0,DM}}{a^{4}}\sqrt{\frac{a^{2}+\beta^{2}}{1+\beta^{2}}} + \rho_{\Lambda} \right)\, ,\label{friedmann}
\\[6pt]
\dot H &=& -\frac{\kappa^{2}}{2}\Big[\rho_b(1+\omega_b) + \rho_r(1+\omega_r) + \rho_{DM}(1+\omega_{DM})\Big]\, ,\label{accel}
\\[6pt]
\dot\rho_b &=& -3H\rho_b(1+\omega_b)\, ,\qquad \dot\rho_r = -3H\rho_r(1+\omega_r)\, ,\label{cons_ord}
\\[4pt]
\dot\rho_{DM} &=& -3H\rho_{DM}(1+\omega_{DM})\, ,
\label{cons_dm}
\end{eqnarray}
where $\kappa^{2}=8\pi G$ and $\rho_{0,i}$ $(i=b,r,DM)$ denote the present-day energy densities of baryons, radiation, and dark matter, respectively. The accelerated expansion of the universe is driven by a cosmological constant $\Lambda$, whose associated energy density is defined as $\rho_\Lambda\equiv\Lambda/\kappa^{2}$. The Hubble parameter is defined as $H \equiv \dot{a}/a$, and an overdot denotes derivative with respect to cosmic time $t$.

We adopt a barotropic equation of state of the form $p_i=\omega_i\rho_i$ for each component. For baryons we set $\omega_b=0$ (pressureless matter) and for radiation $\omega_r=1/3$. For dark matter the equation-of-state parameter is given by Eq.~\eqref{eos_dm}, which once it is substituted into the conservation equation~\eqref{cons_dm} yields the energy density~\eqref{rho_dm}.

An important physical quantity characterizing the relativistic kinetic dark matter fluid is the squared adiabatic sound speed $c_s^2$, which controls the response of the pressure to homogeneous density variations and therefore plays a central role in the evolution of cosmological perturbations.

In the present work, we define the adiabatic sound speed at the background level as
\begin{equation}
c_s^2 \equiv \frac{\delta P}{\delta \rho}\, ,
\end{equation}
which is appropriate for adiabatic perturbations whose local equation of state follows that of the homogeneous background.

We emphasize that, in general, collisionless kinetic systems may exhibit non--adiabatic stress contributions and scale--dependent effective sound speeds at the perturbation level. However, since our analysis focuses on the background evolution and on adiabatic perturbations, the above definition provides a well--defined and physically meaningful characterization of the effective stiffness of the relativistic kinetic gas. In this case, the sound speed is given by
\begin{equation}
    c_s^2(a) \equiv \frac{\delta P}{\delta \rho}
    = \omega(a) + \rho(a)\,\frac{d\omega}{d\rho}\, ,
    \label{Eq:cs2}
\end{equation}
where $\omega(a)=P(a)/\rho(a)$ denotes the equation--of--state parameter of the dark matter component.

Since both $\omega(a)$ and $\rho(a)$ are explicit functions of the scale factor, this definition yields a closed analytic expression for the sound speed,
\begin{equation}
c_s^2(a) = \frac{\beta^2}{3\big(\beta^2 + a^2\big)}\frac{4\beta^2 + 5a^2}{4\beta^2 + 3a^2} \equiv \omega(a)\, f(a)\, ,
\end{equation}
with
\begin{equation}
f(a) \equiv \frac{4\beta^2 + 5a^2}{4\beta^2 + 3a^2}\, .
\label{asymp_beh_f_a}
\end{equation}

From this expression it follows that the sound speed naturally transitions between the relativistic and non--relativistic regimes. At early times, $a \ll \beta$, the dark matter behaves effectively as radiation,
\begin{equation}
c_s^2 \simeq \omega \simeq \frac{1}{3}\, ,
\end{equation}
whereas at late times, $a \gg \beta$, the particles become non--relativistic,
\begin{equation}
c_s^2 \simeq \omega \simeq 0 \, .
\end{equation}

More precisely, the departure from the simple adiabatic relation $c_s^2=\omega$ is controlled by the factor $f(a)$, which varies monotonically with the scale factor as
\begin{equation}
f(a) = \left\lbrace
\begin{matrix}
1 & {\rm for}\quad a \rightarrow 0\, , \\[4pt]
5/3 & {\rm for}\quad a \rightarrow \infty \, .
\end{matrix}
\right.
\end{equation}

Thus, although $c_s^2$ coincides exactly with $\omega$ only in the asymptotic relativistic and non--relativistic limits, their difference remains small throughout the cosmic evolution, ensuring that the sound speed stays of the same order as the equation--of--state parameter at all times.

In order to confront the model with cosmological observations, such as the temperature anisotropies of the \textit{Cosmic Microwave Background} (CMB) and the matter power spectrum (MPS), we consider the linear perturbations of both the spacetime metric and the cosmic fluid components.

The perturbed line element in the Newtonian gauge is written as
\begin{equation}
ds^{2} = a^{2}(\eta)\left[ -(1+2\psi)\, d\eta^{2} + (1-2\phi)\, (dx^2 + dy^2 + dz^2)\right]\, ,
\end{equation}
where $\psi(\vec x,\eta)$ and $\phi(\vec x,\eta)$ are the scalar gravitational potentials, which are expressed as functions of the spatial coordinates $\vec{x}$ and conformal time $\eta$. In the following, we work in Fourier space, where all perturbation variables are decomposed into plane waves and depend on the comoving wave vector $\vec{k}$ and conformal time $\eta$. For notational simplicity, we do not distinguish explicitly between real--space and Fourier--space variables, and the dependence on $\vec{k}$ is understood whenever perturbation equations are written.

In this gauge, after considering the Fourier transform and assuming vanishing anisotropic stress ($\sigma = 0$), the linearized Einstein equations for scalar perturbations read~\cite{Ma:1995ey}
\begin{subequations}
\label{einstein_newtonian}
\begin{align}
k^{2}\phi + 3\mathcal{H} \left(\phi^{\prime} + \mathcal{H}\psi \right) &= -4\pi G a^{2}\,\delta \rho\, ,
\\[1ex]
k^{2}\left(\phi^{\prime} + \mathcal{H}\psi \right) &= 4\pi G a^{2}\,(\bar\rho + \bar P)\, \theta\, ,\label{phitheta}
\\[1ex]
\phi &= \psi\, ,\label{phipsi}
\end{align}
\end{subequations}
where primes denote derivatives with respect to conformal time, and $\mathcal{H}=a'/a$ denotes the conformal Hubble parameter. The quantity $\delta\rho=\sum_i \bar\rho_i \delta_i$ is the total density perturbation, $(\bar\rho+\bar P)\theta=\sum_i(\bar\rho_i+\bar P_i)\theta_i$ is the total momentum density perturbation, and $\theta\equiv \nabla\cdot \vec{v}$ the velocity divergence. Here $\bar{\rho}$ denotes the background energy density, while $\delta\rho$ represents its perturbation.

The evolution of linear perturbations for a generic fluid component, written in terms of the density contrast $\delta\equiv \delta \rho/\bar{\rho}$ and the velocity divergence in Fourier space $\theta \equiv i\vec{k}\cdot\vec{v}$, where $\vec{v}$ denotes the peculiar velocity field of the fluid, is governed by~\cite{Ma:1995ey}
\begin{subequations}
\label{ng_dc_dv}
\begin{eqnarray}
\delta' &=& -(1+\omega)\left(\theta -3\phi'\right) - 3\mathcal{H}\left(c_s^2-\omega\right)\delta\, ,
\\[4pt]
\theta' &=& -\mathcal{H}\left(1-3\omega\right)\theta - \frac{\omega'}{1+\omega}\theta + \frac{c_s^2}{1+\omega}k^2\delta - k^2\sigma + k^2\psi\, .
\end{eqnarray}
\end{subequations}

For the relativistic kinetic gas dark matter component considered here, we assume that the anisotropic stress vanishes, $\sigma = 0$. Although it is a priori not clear whether or not this assumption is consistent with the perturbed Vlasov equation, we take it as reasonable approximation that should work at least for $\beta\ll a$.

Equations~\eqref{einstein_newtonian} together with the fluid perturbation system~\eqref{ng_dc_dv} and the expressions~\eqref{eos_dm} and \eqref{Eq:cs2} for $\omega$ and $c_s^2$ form a closed set that fully determines the coupled evolution of metric and matter perturbations for all cosmological components for the scalar modes, including the relativistic kinetic gas dark matter model studied in this work.

An important cosmological consequence is the impact on the CMB temperature anisotropies. Since the RKG dark matter energy density acquires a relativistic correction for non--vanishing values of $\beta$, the evolution of the anisotropies is modified at early times, leading to a small modification of the effective matter--radiation content prior to recombination.

This induces a time dependence in the gravitational potential $\Phi\equiv \phi = \psi$, which according to Eqs.~\eqref{phitheta} and~\eqref{phipsi} satisfies
\begin{equation}
\Phi' + \mathcal{H}\Phi = -4\pi G a^2 \sum_i (\rho_i + P_i)\, v_i \, ,
\end{equation}
where $v_i$ is the scalar mode $v$ of the peculiar velocity $\vec{v}$ for the $i$-th component, this is, $\vec{v}_i(t,\vec{x}) = \nabla v_i(t,\vec{x})$.

Assuming vanishing anisotropic stress, $\Phi$ satisfies
\begin{equation}
\Phi'' + 3\mathcal{H}\Phi'
+ \left(2\mathcal{H}' + \mathcal{H}^2\right)\Phi
= 4\pi G a^2\,\delta P_{\rm tot}\, .
\end{equation}

In the absence of pressure perturbations, $\delta P_{\rm tot}=0$, and during
matter domination ($a\propto\eta^2$) the equation admits a constant mode,
$\Phi=\mathrm{const}$, plus a rapidly decaying one.

For the RKG dark matter component with $\beta\neq 0$, non--vanishing pressure
perturbations and departures from pure matter domination render $\Phi'\neq 0$
during the radiation--to--matter transition, enhancing the early Integrated
Sachs--Wolfe contribution to the CMB anisotropies.

The temperature anisotropies sourced by the gravitational potentials are given by~\cite{Sachs:1967er}
\begin{equation}
\left(\frac{\Delta T}{T}\right)_{\rm ISW} = 2\int_{\eta_*}^{\eta_0}\Phi'(\eta)\, d\eta \, ,
\end{equation}
so that the early-time evolution of $\Phi$ induced by $\beta$ modifies the driving term of the photon--baryon acoustic oscillations. Here $\eta$ denotes conformal time, $\eta_0$ corresponds to the present epoch, and $\eta_*$ is the conformal time at photon decoupling (last scattering surface). This effect primarily impacts modes entering the horizon before recombination, leading to small but coherent changes in the heights of the acoustic peaks.

In contrast, large angular scales correspond to modes that remain super--horizon
at recombination ($k \ll \mathcal{H}_*$), for which the gravitational potential
remains approximately constant,
\begin{equation}
\Phi(k,\eta) \simeq \text{const.}\, ,
\qquad k \ll \mathcal{H}\, ,
\end{equation}
implying a negligible contribution to the ISW integral. Consequently, the low--$\ell$ CMB anisotropies are largely insensitive to small early-time deviations in the dark sector induced by the RKG model.

In the case of the matter power spectrum, the relativistic kinetic gas model leaves a complementary and physically distinct signature with respect to the standard cold dark matter (CDM) scenario. While the CMB is primarily sensitive to the early--time relativistic behavior of the RKG component through its impact on the gravitational potentials and acoustic oscillations, the matter power spectrum probes the cumulative effect of the residual effective pressure on the growth of dark matter perturbations. As a result, the RKG model induces a characteristic scale--dependent modification of the power spectrum, with a suppression of small--scale power that reflects the transient relativistic nature of the dark sector and is controlled by the parameter~$\beta$.

Once the RKG evolves with $\omega=0$ during matter domination, the second--order equation for $\delta_{\rm RKG}$ arising from the combination of Eqs.~\eqref{ng_dc_dv} yields,
\begin{equation}
\delta''+\mathcal{H}\delta'+\left(c_s^2 k^2 - 4\pi G a^2 \rho\right)\delta \simeq 0 \, ,
\end{equation}
which explicitly shows the competition between gravitational clustering and the
pressure--induced term proportional to $c_s^2 k^2$.

Recalling that $4\pi G a^2\rho\propto \mathcal{H}^2$, modes satisfying
\begin{equation}
k \gg k_J(a) \equiv \frac{\mathcal{H}}{c_s}\, 
\end{equation}
experience pressure support and suppressed growth, while large--scale modes with $k \ll k_J$ grow in the standard cold dark matter fashion. Since the sound speed $c_s(a)$ decreases with time, modes entering the horizon later are less affected by this suppression. As a result, the late--time matter power spectrum in the RKG model exhibits a scale--dependent suppression relative to its cold dark matter counterpart, i.e.,
\begin{equation}
P_{\rm RKG}(k) \simeq T^2_{\rm RKG}(k;\beta)\, P_{\rm CDM}(k)\, ,
\end{equation}
where the transfer function $T_{\rm RKG}(k;\beta)$ exhibits a smooth cutoff at small scales. This behavior is qualitatively analogous to that induced by warm dark matter or massive neutrinos, although the precise scale and shape of the suppression in our model are controlled by the parameter $\beta$ and reflect the transient relativistic nature of the RKG component.

\section{Numerical solutions}\label{num_sols}

\begin{figure}[h!]
    \centering
    \includegraphics[width=12cm]{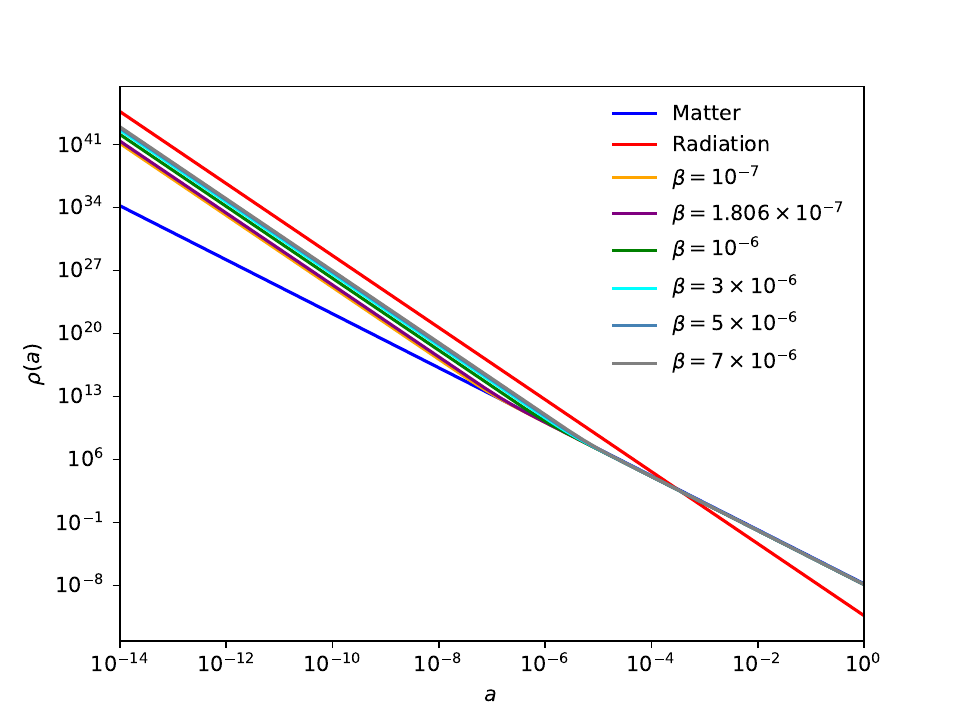}
    \caption{Cosmological evolution of the energy density $\rho_{DM}(a)$ for several values of the relativistic parameter $\beta$. The blue and red curves represent, respectively, the CDM and radiation components, while the colored lines show the kinetic--gas dark matter model for increasing values of $\beta$. All curves are normalized to the present-day abundances.}
    \label{densities}
\end{figure}

Figure~\ref{densities} shows the background evolution of the dark--matter energy
density obtained from a modified version of the Boltzmann code \textsc{CLASS}, where we implemented the relativistic kinetic gas dark--matter model. All curves correspond to the mono--energetic DF and different values of the effective velocity parameter~$\beta$. For comparison, the standard CDM and radiation evolutions are also shown.

The numerical solutions clearly exhibit the expected transitory behavior: for any fixed value of $\beta$, the kinetic gas behaves as a radiation--like component at early times and evolves smoothly towards a dust--like behavior at late times. As anticipated by the analytical expression for the energy density, the transition occurs around the scale factor
\begin{equation}
    a_{\mathrm{tr}}\sim \beta \, ,
\end{equation}
so that larger values of $\beta$ delay the radiation--to--matter transition,
while smaller $\beta$ induce an earlier convergence to the CDM scaling $\rho\propto a^{-3}$.

In the numerical scans performed with \textsc{CLASS} we fixed the present dark--matter abundance $\Omega_{0,DM}$ to its observed value and varied $\beta$. For sufficiently small $\beta$ the background evolution is indistinguishable from CDM well before the epoch of matter--radiation equality, whereas for larger $\beta$ the extended relativistic phase yields significant deviations.

\begin{figure}[h!]
    \centering
    \includegraphics[width=12cm]{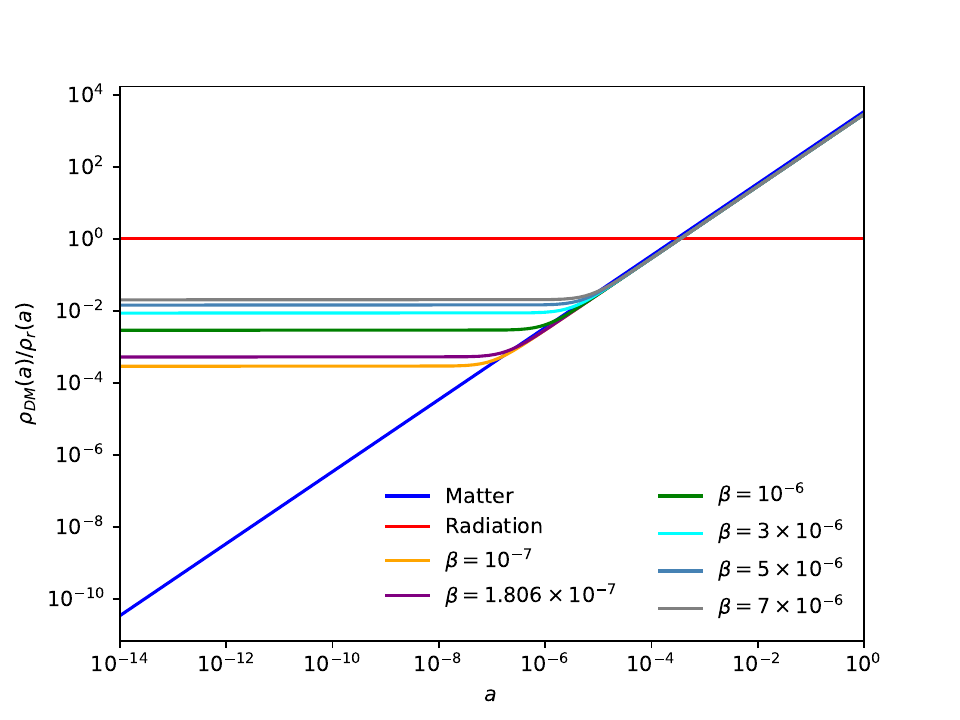}
    \caption{Evolution of the ratio $\rho_{DM}(a)/\rho_r(a)$ for several values of the relativistic velocity parameter $\beta$. The blue line shows the standard cold matter scaling $\rho_m/\rho_r \propto a$, while the red line corresponds to pure radiation behavior $\rho_r/\rho_r = 1$.}
    \label{densities_comp}
\end{figure}

Figure~\ref{densities_comp} shows the cosmological evolution of the ratio between the energy density of the relativistic kinetic gas dark matter component and that of standard radiation, $\rho_{DM}(a)/\rho_r(a)$, for several representative values of the velocity parameter $\beta$. For reference, we also display the standard behaviors corresponding to cold matter, $\rho_m/\rho_r$, shown as the blue rising line, and to pure radiation shown as the horizontal red line.

At very early times, when $a \ll \beta$, the kinetic gas is effectively relativistic and its energy density scales as $\rho_{DM}\propto a^{-4}$. As a result, the ratio $\rho_{DM}/\rho_r$ remains approximately constant, forming the plateau observed in Fig.~\ref{densities_comp}. The amplitude of this early--time plateau depends on the value of $\beta$, with smaller velocities producing a smaller relativistic contribution to the total radiation budget.

As the Universe expands and the scale factor approaches the transition epoch
$a \sim \beta$, the particles gradually become non--relativistic. Once this
transition is completed, the energy density redshifts as
\begin{equation}
\rho_{DM} \propto a^{-3}\, ,
\end{equation}
and consequently
\begin{equation}
\frac{\rho_{DM}}{\rho_r} \propto a \, ,
\end{equation}
matching the behavior of cold matter. This explains why all curves converge at
late times to the same linear slope represented by the reference matter line.
Larger values of $\beta$ delay the onset of the non--relativistic regime,
effectively extending the period in which the kinetic gas behaves as radiation,
whereas smaller values of $\beta$ shift the transition to earlier epochs.

This phenomenology closely resembles that of massive neutrinos. In the early
Universe, neutrinos behave as relativistic species and contribute as an additional radiation component, yielding a constant ratio $\rho_{\nu}/\rho_r$. As the cosmic temperature drops below the neutrino mass scale, neutrinos transition to the non--relativistic regime and start redshifting as matter, leading to $\rho_{\nu}/\rho_r \propto a$. The relativistic kinetic gas dark matter model provides an analogous behavior, with the single velocity parameter $\beta$ controlling the location of the transition in a manner similar to how the neutrino mass determines the neutrino transition epoch.

The crucial difference lies in the present--day abundance: while massive neutrinos represent only a subdominant fraction of the matter density, the kinetic gas species accounts for the entire dark matter component. This imposes strong constraints on its early relativistic contribution. Values of $\beta$ that are too large would imply an excessive radiation--like energy density at early times, in conflict with observational bounds on the effective number of relativistic species and the standard expansion history. For this reason, consistency with the background cosmological evolution leads to an upper bound on the allowed values of $\beta$, as discussed in the previous section. See Sec.~\ref{sec:Neff_constraint} for a detailed discussion and BBN constraints.

Figure~\ref{densities_comp} thus illustrates how the relativistic kinetic gas dark matter naturally interpolates between radiation--like behavior at early times and standard cold dark matter dynamics at late times, while remaining fully consistent with the observed cosmological background evolution.

\begin{figure}[h!]
    \centering
    \includegraphics[width=12cm]{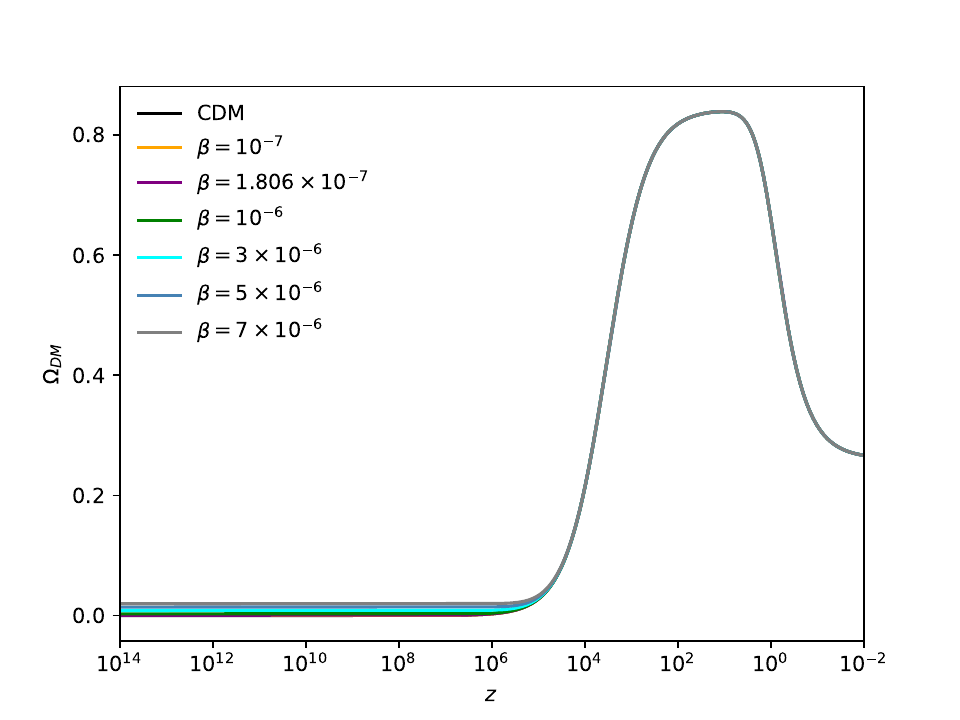}
    \caption{Cosmological evolution of the dark matter density parameter $\Omega_{DM}(z)$ for several values of $\beta$. The standard CDM evolution (black curve) is shown for comparison.}
    \label{Omegas}
\end{figure}

In order to explore the impact of the velocity parameter $\beta$ on the background evolution of the dark matter component, we show in Fig.~\ref{Omegas} the cosmological evolution of the dark matter density parameter $\Omega_{DM}(z)$ for several representative values of $\beta$, spanning nearly two orders of magnitude around the lower bound obtained in the previous section. The standard cold dark matter (CDM) case is included for reference.

We recall that all curves are normalized to the same present-day abundance,
$\Omega_{0,DM}=0.26$. As can be seen from Fig.~\ref{Omegas}, the overall evolution of $\Omega_{DM}$ remains practically indistinguishable from the CDM prediction for the entire range of explored $\beta$ values. Any dependence on $\beta$ is restricted to the very early radiation-dominated era, at redshifts
$z\gtrsim 10^{6}$, where the relativistic kinetic gas behaves effectively as an
additional radiation-like component before its equation of state relaxes towards the non-relativistic regime. 

\begin{figure}[h!]
    \centering
    \includegraphics[width=12cm]{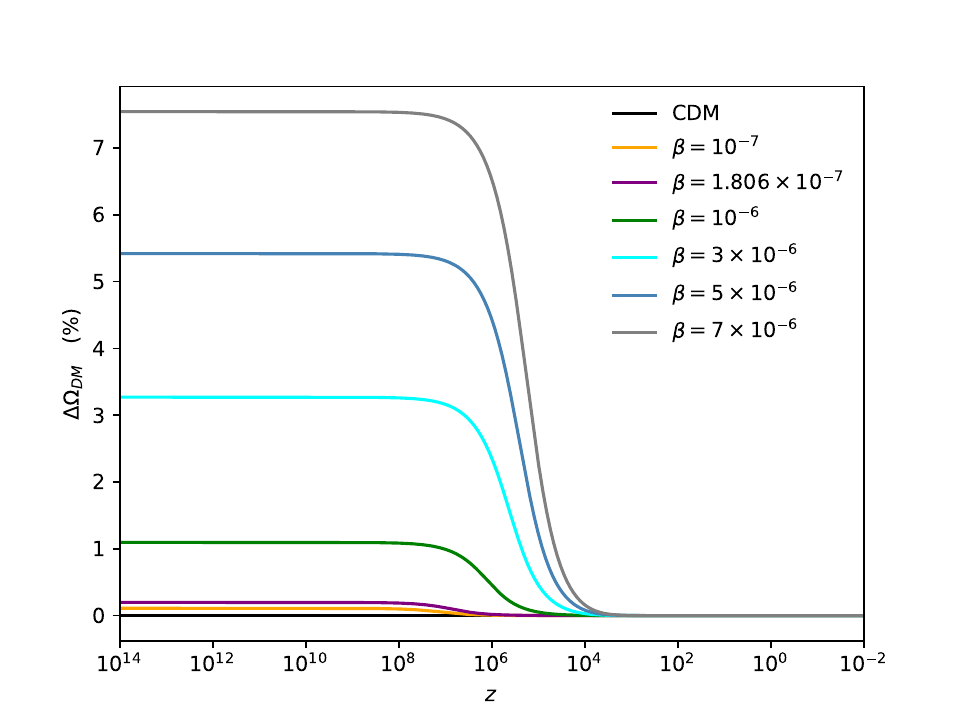}
    \caption{Relative deviation of the dark matter density parameter with respect to CDM, $\Delta\Omega_{DM}$, for the same set of $\beta$ values shown in Fig.~\ref{Omegas}.}
    \label{Omegas_diff}
\end{figure}

The relative deviation of $\Omega_{DM}$ with respect to the CDM case is quantified in Fig.~\ref{Omegas_diff}, where we plot $\Delta\Omega_{DM}\equiv
|\Omega_{DM}-\Omega_{CDM}|/\Omega_{CDM}$ in percentage units.

We observe that the maximal deviations occur at very early times and remain small for all considered values of $\beta$, rendering the background expansion essentially indistinguishable from the standard CDM cosmology. As $\beta$ increases, corresponding to larger present-day velocity dispersions, the duration of the radiation-like regime becomes longer, producing a larger temporary suppression of $\Omega_{DM}$ before the relativistic-to nonrelativistic transition is completed. Nevertheless, even for the largest values shown ($\beta \sim 7\times 10^{-6}$), the resulting departures remain moderate and confined to epochs far earlier than those directly accessible to late-time cosmological observations.

This behavior is qualitatively analogous to that expected for massive neutrinos, whose background energy densities also interpolate between radiation-like and matter-like scaling regimes. In both cases, the transition from
$\rho \propto a^{-4}$ to $\rho \propto a^{-3}$ induces a transient modification of the total matter abundance at early times. However, in the present model, the mono--energetic nature of the DF fixes the transition scale uniquely through the single parameter $\beta$, in contrast with the thermal distribution used for neutrinos. Moreover, once the kinetic gas becomes non-relativistic, its subsequent evolution exactly follows that of standard cold dark matter, ensuring convergence to the CDM background dynamics by $z \lesssim 10^{4}$ for all viable values of $\beta$.

\begin{figure}[h!]
    \centering
    \includegraphics[width=12cm]{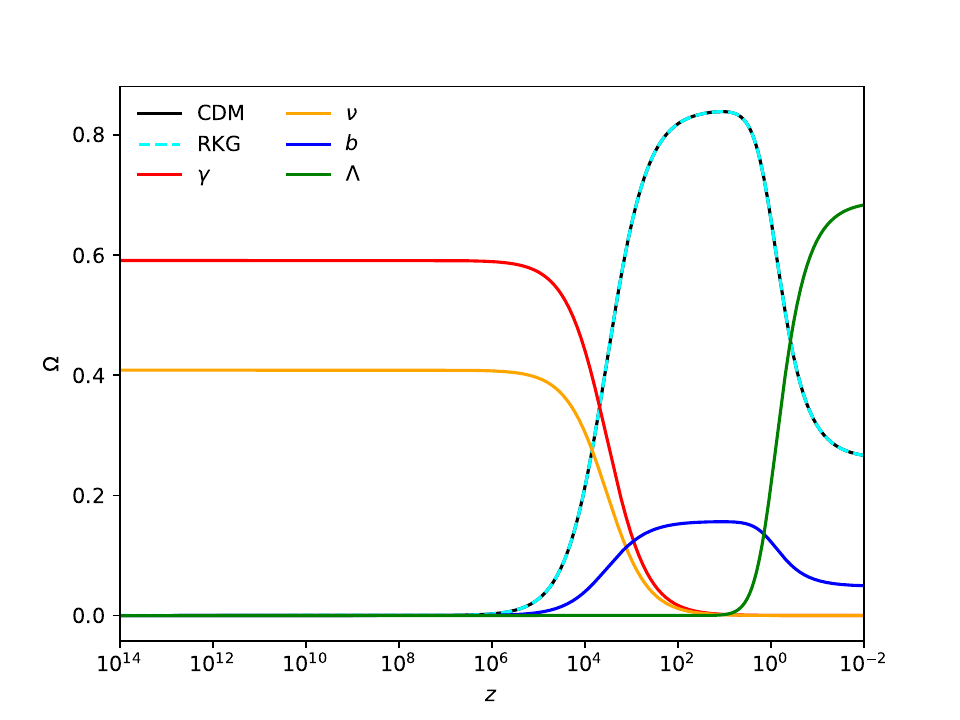}
    \caption{Cosmological evolution of the density parameters of all relevant
    components. The standard CDM case is shown as the black reference curve, while the RKG model is represented by the dashed cyan curve.}
    \label{Omegas_all_new}
\end{figure}

We verify that the introduction of the relativistic kinetic gas dark matter component preserves the standard background evolution of the remaining cosmic fluids. In Fig.~\ref{Omegas_all_new} we display the cosmological evolution of the density parameters of all relevant components: photons ($\gamma$, red curve), neutrinos ($\nu$, orange curve), baryons ($b$, blue curve), the cosmological constant ($\Lambda$, green curve), and the relativistic kinetic gas dark matter (cyan dashed curve), together with the CDM reference case (black curve).

As the figure clearly shows, the background evolution of photons, neutrinos,
baryons, and dark energy is essentially indistinguishable from that of the
standard $\Lambda$CDM cosmology. The relativistic kinetic gas dark matter
component closely tracks the CDM behavior once it becomes non--relativistic,
ensuring that the matter--dominated epoch and the onset of late--time cosmic
acceleration proceed exactly as in the concordance model.

At early times, when $a \ll \beta$ (equivalently $z \gg \beta^{-1}$), the relativistic kinetic gas dark matter component behaves effectively as an additional radiation species, contributing to the total relativistic energy budget alongside photons and neutrinos. As the Universe expands and the scale factor approaches $a \sim \beta$, the equation of state of the kinetic gas smoothly transitions from the radiation-like regime ($\omega=1/3$) to the pressureless matter regime ($\omega\simeq 0$), and the dark matter density parameter rises accordingly to join the baryonic matter component in driving the subsequent matter-dominated era.

Importantly, for all phenomenologically viable values of $\beta$ identified in
our analysis, this early-time relativistic phase produces only negligible
modifications to the global expansion history.  Consequently, the standard
sequence of cosmological epochs --- radiation domination, matter domination,
and late-time accelerated expansion driven by $\Lambda$ --- remains fully
preserved.

\begin{figure}[h!]
    \centering
    \includegraphics[width=12cm]{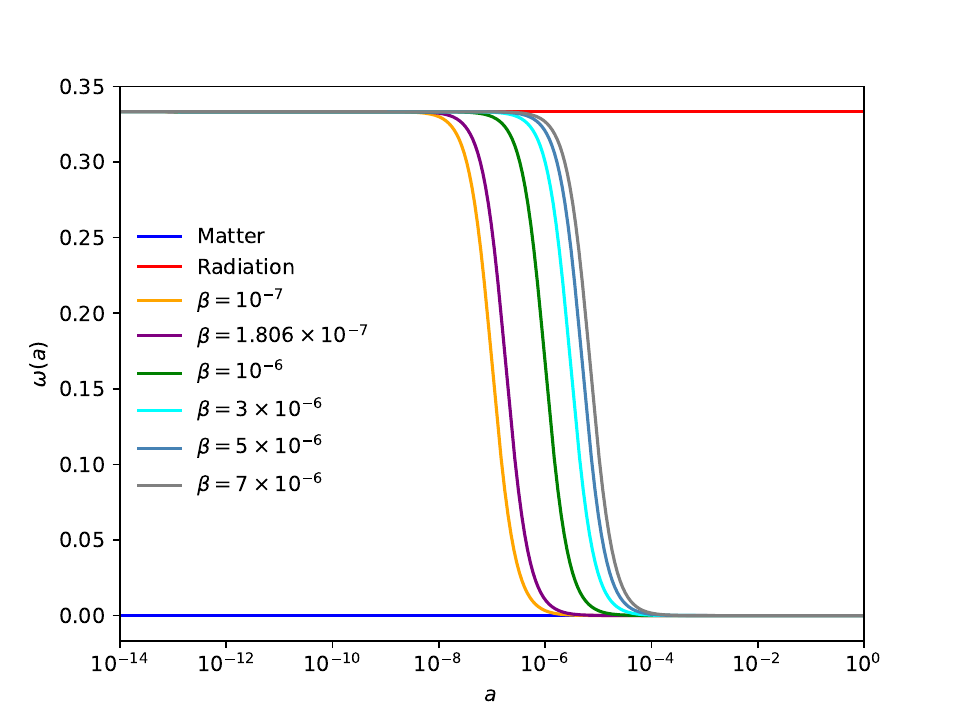}
    \caption{Evolution of the equation--of--state parameter $\omega_{DM}(a)$ of the relativistic kinetic gas dark matter for representative values of $\beta$. The horizontal red line denotes the radiation value $\omega_r=1/3$, while the blue line indicates the cold dark matter limit $\omega_{\rm cdm}=0$.}
    \label{eos_new}
\end{figure}

We now turn to the evolution of the effective equation--of--state parameter
of the relativistic kinetic gas dark matter. As anticipated in Eq.~\eqref{eos_dm}, the model exhibits a clear asymptotic behavior controlled by the velocity parameter $\beta$. In Fig.~\ref{eos_new} we show the evolution of $\omega_{DM}(a)$ for the same representative values of $\beta$ used in the
previous sections.

At early times, when $a \ll \beta$, the kinetic gas remains in the relativistic
regime and behaves as an additional radiation component, with
$\omega_{DM}\simeq 1/3$. As the Universe expands and the scale factor approaches
$a \sim \beta$, particle momenta redshift and the equation of state departs from
the relativistic value, marking the onset of the transition toward
matter--like behavior. In the late--time regime $a \gg \beta$, the particles are
effectively non--relativistic and the equation of state asymptotically approaches the cold dark matter limit $\omega_{DM}\simeq 0$.

The parameter $\beta$ governs both the timing and the duration of the transition. Since the transition occurs roughly at $a_{\rm tr}\sim\beta$, \emph{smaller} values of $\beta$ produce an earlier transition to the non--relativistic regime (i.e. at smaller scale factor or larger redshift), so the dark matter component behaves as CDM for a longer subsequent period. Conversely,
\emph{larger} values of $\beta$ postpone the transition, so the kinetic gas remains radiation--like for an extended cosmic time. This behavior is illustrated in Fig.~\ref{eos_new}: curves with smaller $\beta$ depart earlier from the radiation plateau and converge sooner to $\omega_{\rm cdm}=0$, while curves with larger $\beta$ stay close to $\omega=1/3$ over a longer redshift interval.

This smooth interpolation between relativistic and non--relativistic dynamics is
a hallmark of the kinetic gas description and is analogous, at the background
level, to the transition exhibited by massive neutrinos. However, unlike the
thermal neutrino case, where the transition is driven by the continuous momentum
distribution and ongoing decoupling dynamics, here it is entirely controlled by
the single parameter $\beta$ associated with the mono--energetic initial
configuration of the dark matter particles.

These results demonstrate that the relativistic kinetic gas model is fully
consistent with the standard cosmological background expansion, with deviations
confined to the deep radiation era and controlled by the single parameter
$\beta$. This robustness naturally motivates an examination of the model at the
perturbative level, where even small early-time departures from the standard
evolution may be amplified and leave observable imprints on cosmological
observables such as the CMB anisotropies and the matter power spectrum.

\begin{figure}[h!]
    \centering
    \subfloat[{CMB temperature anisotropies.}]{\includegraphics[width=7cm]{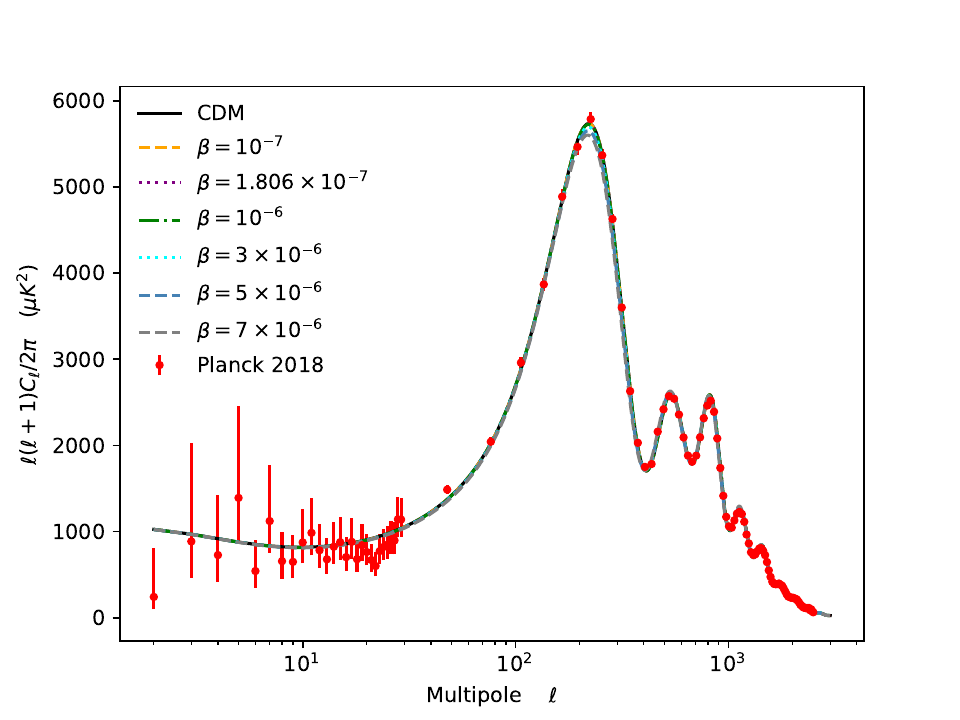}}
    \qquad
    \subfloat[{Relative difference $\Delta D_{\ell}$ for the same set of $\beta$ values.}]{\includegraphics[width=7cm]{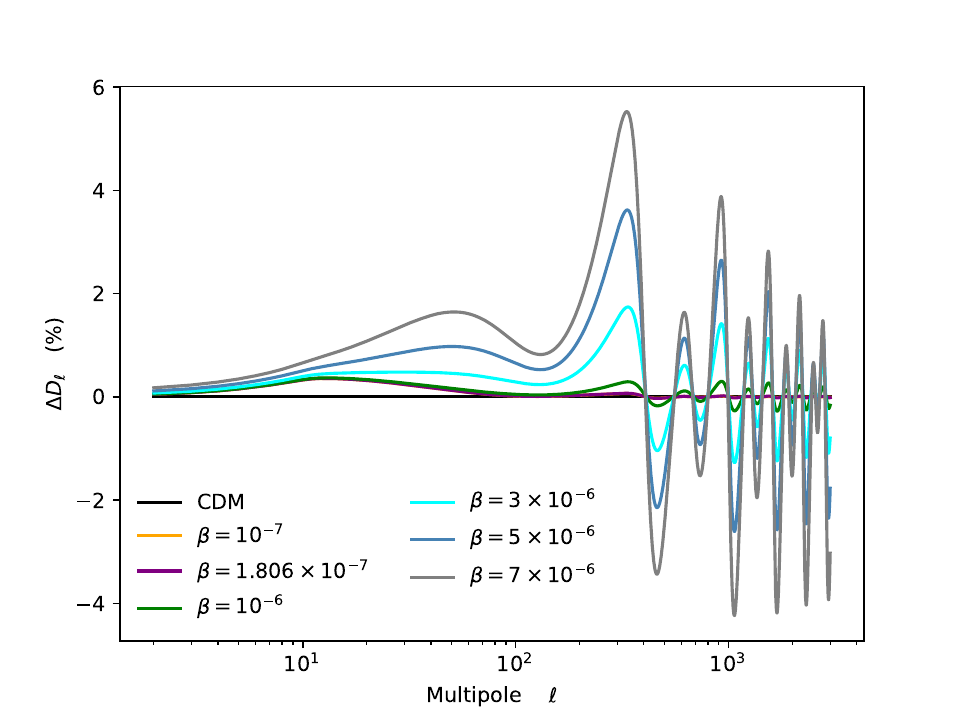}}
    \caption{Left panel: CMB temperature anisotropy spectra computed numerically for various values of $\beta$. Right panel: corresponding relative differences with respect to the standard CDM scenario, illustrating the sensitivity of the high-$\ell$ tail to the RKG model.}
    \label{fig:cmb_rkg}
\end{figure}

\begin{figure}[h!]
    \centering
    \subfloat[{Matter power spectrum $P(k)$.}]{\includegraphics[width=7cm]{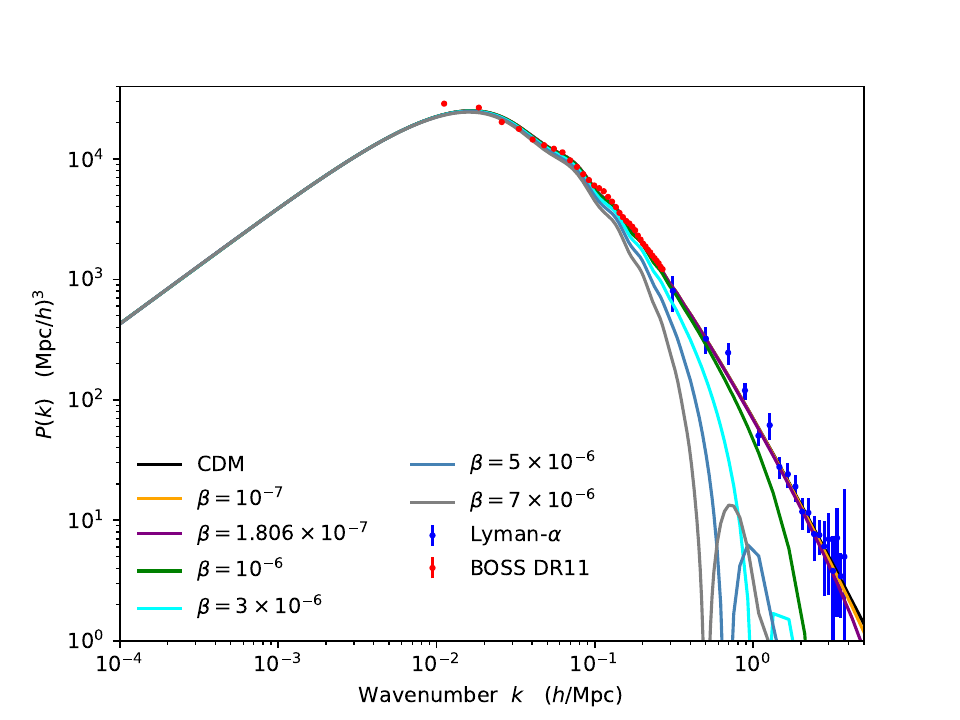}}
    \qquad
    \subfloat[{Relative difference $\Delta P(k)$, showing the scale-dependent suppression induced by $\beta$.}]{\includegraphics[width=7cm]{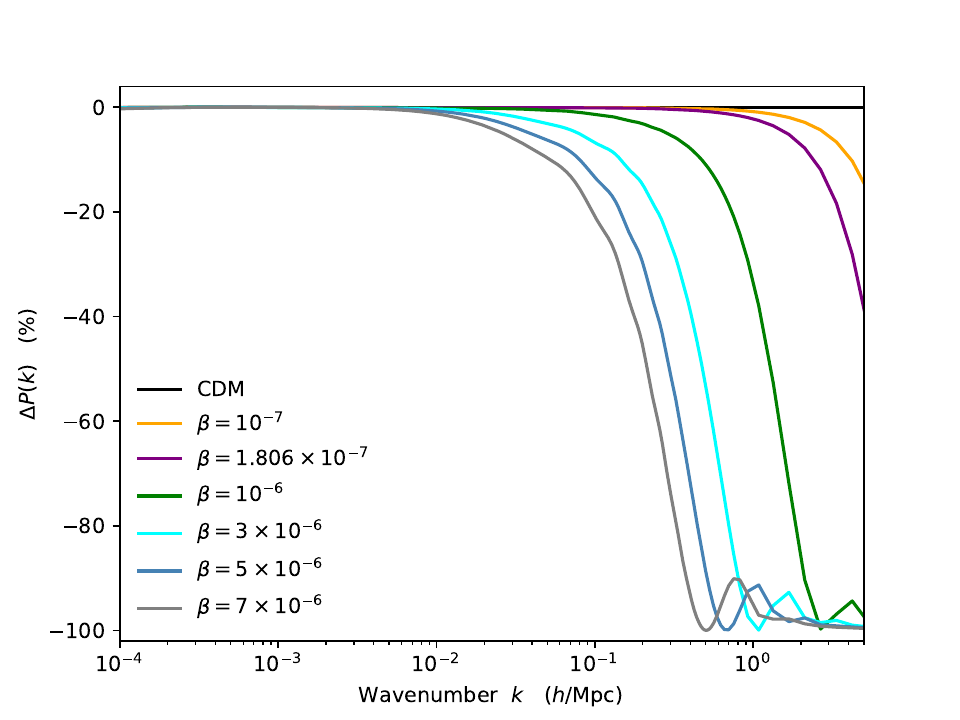}}
    \caption{Left panel: numerical predictions for the matter power spectrum including RKG interactions, compared with CDM and observational data. Black corresponds to CDM, colored lines to increasing $\beta$. Blue and red points indicate Lyman-$\alpha$ forest data~\cite{Chabanier:2019eai} and BOSS DR11 measurements \cite{BOSS:2013rlg}, respectively. Right panel: relative difference with respect to CDM, emphasizing the suppression of small-scale structure as $\beta$ increases.}
    \label{fig:pk_rkg}
\end{figure}

In this context, Figures~\ref{fig:cmb_rkg} and~\ref{fig:pk_rkg} illustrate the
impact of the RKG model, as parameterized by $\beta$, on these key observables,
namely the CMB temperature anisotropies and the matter power spectrum. Together,
these results elucidate how subtle deviations from the standard cold dark matter
scenario propagate into both the linear and quasi-linear regimes of structure
formation.

In the CMB temperature anisotropy spectra (Fig.~\ref{fig:cmb_rkg}a), the black solid line represents the standard CDM prediction, while the colored lines correspond to increasing values of $\beta$. For $\beta \lesssim 10^{-6}$, the spectra remain nearly identical to the CDM case across all multipoles, indicating that the background evolution and linear perturbations are only marginally affected. As $\beta$ increases, however, deviations appear primarily at high multipoles ($\ell \gtrsim 500$), corresponding to angular scales smaller than a degree. The relative difference $\Delta D_\ell$ (Fig.~\ref{fig:cmb_rkg}b) highlights that these deviations are oscillatory in nature, reflecting modifications in the acoustic oscillation amplitude and phase induced by the RKG model. As mentioned in the previous section, this behavior arises because a larger $\beta$ slightly alters the effective matter content at early times, modifying the gravitational potentials and thus the photon-baryon acoustic oscillations before recombination. The low-$\ell$ modes remain essentially unchanged, consistent with the fact that the integrated Sachs-Wolfe effect and large-angle anisotropies are less sensitive to small deviations in the dark sector at early times.

The impact on the matter power spectrum $P(k)$ is even more pronounced at small scales. Figure~\ref{fig:pk_rkg}a shows that for increasing $\beta$, there is a scale-dependent suppression of power, most evident at $k \gtrsim 0.1 \ h/$Mpc, while large-scale modes ($k \lesssim 0.01 \ h/$Mpc) are largely unaffected. The corresponding relative differences $\Delta P(k)$ (Fig.~\ref{fig:pk_rkg}b) quantify this suppression, which reaches nearly $100\%$ for the largest $\beta$ considered. This scale-dependent behavior is due to the fact that the RKG dark matter model introduces a small effective pressure or damping in the dark sector, which suppresses the growth of small-scale density perturbations while leaving large-scale modes, which entered the horizon later, essentially unaltered. Consequently, the shape of $P(k)$ is modified in a manner analogous to warm dark matter (WDM) or massive neutrinos, albeit with a distinct parametric dependence on $\beta$.

Observational data provide critical constraints on this behavior. The blue and red points in Fig.~\ref{fig:pk_rkg}a correspond to Lyman-$\alpha$ forest \cite{Chabanier:2019eai} and BOSS DR11 \cite{BOSS:2013rlg} measurements, which probe small-scale structure at redshifts $z\sim 2-5$. Only models with $\beta \lesssim 10^{-6}$ remain compatible with these data, as larger values produce excessive suppression of small-scale power. This result is consistent with previous findings using Planck 2018 CMB data, where small $\beta$ values are required to preserve the observed acoustic peak structure.

In summary, Figs.~\ref{fig:cmb_rkg} and~\ref{fig:pk_rkg} demonstrate that the RKG parameter $\beta$ predominantly affects high-$\ell$ CMB modes and small-scale matter fluctuations. The effects are subtle for $\beta \lesssim 10^{-6}$, ensuring compatibility with current cosmological observations, but grow rapidly for larger values. Thus, for small values of $\beta$ the RKG model behave as a CDM-like fluid at both background and linear perturbation levels, while larger couplings introduce scale-dependent modifications that are constrained by both the CMB and the Lyman-$\alpha$ forest.

To conclude this section, we note that all figures presented above include the reference value $\beta = 1.806\times 10^{-7}$, which has not been explicitly motivated so far. Since $\beta \simeq \beta_0 \equiv |\vec v_0|/c$, this choice corresponds to a present--day particle velocity $v_0 \simeq 54~\mathrm{m/s}$. This value is physically well motivated, as it coincides with the upper bound on the root--mean--square dark matter velocity reported in~\cite{Armendariz-Picon:2013jej}, $v_{\rm rms} \leq 54~\mathrm{m/s}$, where the authors base their analysis on a thermal DF.

Importantly, this reference value does not correspond to a best--fit parameter inferred from the data, but rather serves as a conservative benchmark illustrating the regime in which the relativistic kinetic gas model remains fully consistent with the standard cosmological background evolution. As shown above, for $\beta \sim \mathcal{O}(10^{-7})$ the deviations from $\Lambda$CDM remain at the sub--permille level throughout cosmic history. In the following statistical analysis, we quantify how observational constraints translate into upper bounds on $\beta$, and assess whether physically motivated values such as this benchmark lie well within the allowed parameter space.

\section{Statistical constraints}
\label{sec:statistics}

In our analysis we employ a combination of CMB and large–scale structure likelihoods to tightly constrain the cosmological parameters of the RKG model. For the CMB, we use the Planck 2018 temperature and polarization likelihoods\footnote{The Planck 2018 CMB likelihoods used in this work are publicly available at~\url{http://pla.esac.esa.int/pla/\#cosmology}.}, including the high–multipole $TT$, $TE$, and $EE$ power spectra and the low–multipole temperature and polarization likelihoods~\cite{Planck:2019nip}, which together capture the acoustic peak structure and reionization signal in the angular power spectra of the CMB. In addition, we include the Planck CMB lensing likelihood, based on the reconstruction of the CMB lensing potential power spectrum, which provides independent information on the growth of structure and helps break parameter degeneracies~\cite{Planck:2018lbu}.

To complement the high–redshift information from the CMB, we incorporate baryon acoustic oscillation (BAO) measurements at low redshift. Specifically, we include the consensus BAO likelihood from the final data release of the Baryon Oscillation Spectroscopic Survey (BOSS DR12), which combines anisotropic measurements of the BAO scale from galaxy clustering at effective redshifts $z\simeq0.38$, $0.51$, and $0.61$~\cite{BOSS:2016wmc}. We also use a compilation of low–redshift BAO measurements from a variety of surveys, including the 6dF Galaxy Survey and other small–$z$ BAO datasets, which provide additional distance constraints at $z\lesssim0.3$ and help anchor the late–time expansion history~\cite{Beutler:2011hx,Ross:2014qpa}. The combination of Planck and BAO data, which will be denoted as Planck 2018, thus constrains both the early–Universe physics encoded in the CMB and the late–time geometry and expansion history of the Universe.

Table~\ref{tab:parameters} summarizes the cosmological parameters employed in our analysis. For each parameter, we report the mean value, the minimum and maximum priors considered in the sampling (where applicable), the step size $\sigma$, the scaling factor used in the sampler, and the type of parameter. All parameters listed in the table correspond to fundamental cosmological quantities that were directly sampled using the \textsc{Monte Python} parameter sampler~\cite{Audren:2012wb,Brinckmann:2018cvx}.
\begin{table}[h!]
\centering
\caption{Parameters used in the Monte Python runs. For each parameter, we report the mean value, the prior minimum and maximum values, and the scaling used in the sampler. Parameters $YHe$, $H_0$, and $\sigma_8$ were treated as derived quantities and are not listed here.}
\label{tab:parameters}

\begin{tabular}{lcccc}
\hline\hline
Parameter & Mean & Prior min & Prior max & Scale \\
\hline
$\omega_b = \Omega_b h^2 $        
& 2.2377 
& --- 
& --- 
& 0.01 \\

$100\,\theta_s$  
& 1.04110 
& --- 
& --- 
& 1 \\

$\ln(10^{10}A_s)$ 
& 3.0447 
& --- 
& --- 
& 1 \\

$n_s$          
& 0.9659 
& --- 
& --- 
& 1 \\

$\tau_\mathrm{reio}$     
& 0.0543 
& 0.004 
& --- 
& 1 \\

$\log \beta$ 
& -7 
& -10 
& -2 
& 1 \\

$\omega_\mathrm{rkg} = \Omega_{rkg}h^2$    
& 0.12011 
& 0.10 
& 0.14 
& 1 \\
\hline\hline
\end{tabular}
\end{table}

In addition, several parameters, namely $YHe$, $H_0$, and $\sigma_8$, were treated as derived quantities and are not included in this table. Although they are not directly sampled in our Monte Python runs, these quantities are particularly relevant in the context of our RKG dark matter model. The helium fraction $YHe$ influences the recombination history and thus the detailed shape of the CMB anisotropies, which are sensitive to the relativistic behavior of the dark matter component at early times. The Hubble parameter $H_0$ sets the overall expansion rate, affecting the transition of the RKG from a relativistic to a non-relativistic regime and the corresponding impact on distance measurements. Finally, $\sigma_8$ quantifies the amplitude of matter fluctuations, which in our model is modified by the kinetic pressure of the dark matter particles, altering the growth of large-scale structure.  Hence, accurately computing these derived parameters is essential for interpreting how the primary sampled parameters translate into observable predictions within the RKG framework and for robust comparison with cosmological data.

\begin{figure}[h!]
    \centering
    \includegraphics[width=12cm]{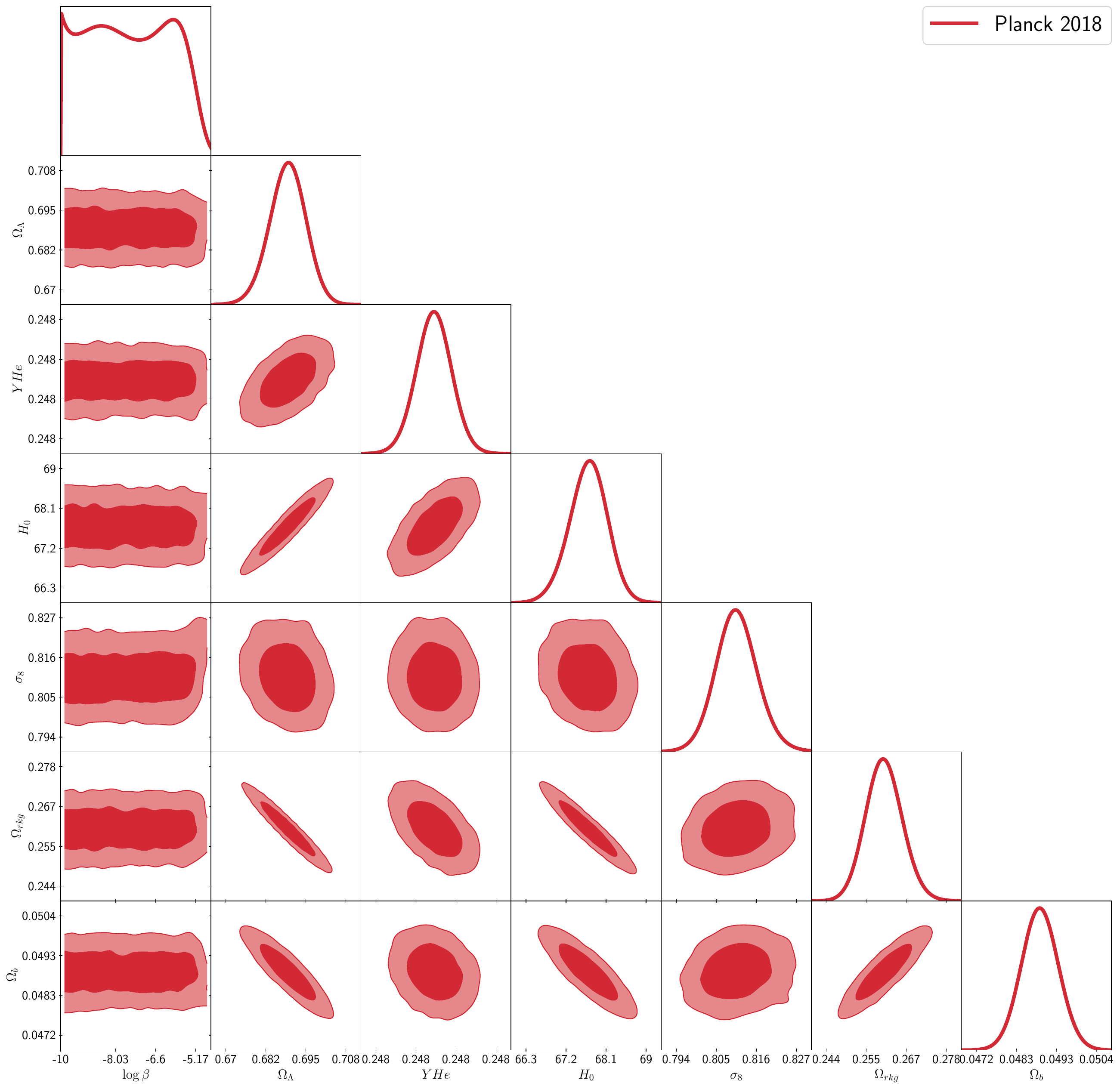}
    \caption{Marginalized posterior distributions for the relativistic kinetic gas (RKG) dark matter model using Planck 2018 data. The $68\%$ and $95\%$ credible regions are shown for $\{\Omega_b, \Omega_{\rm rkg}, \log\beta, H_0, \Omega_\Lambda, YHe, \sigma_8\}$. See text for more details.}  
    \label{2d_posteriors}
\end{figure}

Figure~\ref{2d_posteriors} shows the marginalized posterior distributions and joint confidence regions obtained from the Planck 2018 data set. Our analysis primarily focuses on the set of parameters $\{\Omega_b, \Omega_{\rm rkg}, \log\beta, H_0, \Omega_\Lambda, YHe, \sigma_8\}$. This choice reflects the fact that the standard cosmological parameters $\{100\,\theta_s, \ln(10^{10}A_s), n_s, \tau_\mathrm{reio}\}$ remain essentially unchanged relative to $\Lambda$CDM, as they are largely determined by inflationary physics and the early-universe conditions. In contrast, $\Omega_{\rm rkg}$ and $\log \beta$ directly encode the properties of the RKG model component, such as its present-day density and relativistic-to-non--relativistic transition. Hence, these parameters are the ones most sensitive to the modifications introduced by the RKG model and are crucial for interpreting its impact on observables like the CMB and large-scale structure. 

The marginalized distributions exhibit well--behaved, approximately Gaussian posteriors for the standard cosmological parameters. In particular, the baryon density $\Omega_b$ and helium fraction $YHe$ remain tightly constrained around their standard values, indicating that the introduction of the RKG component does not significantly alter early--time physics such as BBN or recombination. Similarly, the inferred value of $\Omega_\Lambda$ is consistent with the $\Lambda$CDM expectation, confirming that the background expansion history remains largely unaffected.

The kinetic parameter $\log \beta$ exhibits a posterior distribution characterized by a plateau at low values and a sharp decline above a certain threshold. This behavior reflects that Planck 2018 data are fully consistent with the CDM limit at very small values of $\log \beta$ (i.e., $\beta \to 0$), while effectively constraining larger values that would induce significant relativistic effects in the dark matter component. Consequently, only an upper limit on $\log \beta$ can be robustly reported. The constraint on $\Omega_{\rm rkg}$ remains compatible with the standard cold dark matter abundance, showing that the RKG scenario can reproduce the observed matter density while allowing for a non-negligible relativistic contribution at early times.

The two-dimensional posterior distribution between $\log \beta$ and $\Omega_{\rm rkg}$ exhibits nearly horizontal contours. This indicates that $\Omega_{\rm rkg}$ is well constrained by the data, while $\log \beta$ remains effectively unconstrained at low values, extending across the full sampled range. Only at sufficiently high values does the posterior decline sharply, reflecting the upper limit imposed by Planck 2018 data. Consequently, there is no significant degeneracy between these two parameters: the dark matter abundance is determined independently of the kinetic parameter, and the data are fully compatible with the CDM limit corresponding to very small $\log \beta$.

The marginalized posterior distributions do not exhibit any statistically significant degeneracy between $\Omega_{\rm rkg}$ and $\sigma_8$, indicating that the constraints on the amplitude of matter clustering are largely determined by the standard cosmological parameters and are not strongly affected by the kinetic properties of the dark matter component. Conversely, the marginalized posteriors show no evidence for strong degeneracies between $\log\beta$ and $\Omega_b$ or $YHe$, suggesting that, at the background and linear perturbation level, the kinetic dark matter sector remains largely decoupled from baryonic physics within the precision of Planck 2018 data.

The Hubble parameter $H_0$ shows only weak correlations with the RKG parameters, suggesting that this model does not introduce significant shifts in the late--time expansion rate. As a result, the RKG framework neither exacerbates nor resolves the existing Hubble tension, but remains fully compatible with Planck measurements.

Table~\ref{tab:rkg_constraints} summarizes the cosmological constraints obtained for the RKG model using Planck 2018 CMB data. For the standard cosmological parameters, we report best--fit values, marginalized posterior means with $68\%$ credible intervals, and $95\%$ bounds whenever available, providing a complete characterization of their uncertainties under the assumptions of our model. 
\begin{table}[h!]
\centering
\caption{Constraints on the RKG model from Planck 2018 CMB data. Quoted uncertainties correspond to $68\%$ credible intervals, while the last two columns show the $95\%$ bounds when available. The minimum of the negative log--likelihood and the corresponding chi--squared value are $-\ln \mathcal{L}_{\rm min} = 1392.4$ and $\chi^2_{\rm min} = 2785$, respectively.}
\label{tab:rkg_constraints}

\begin{tabular}{lcccc}
\hline\hline
Parameter & Best--fit & Mean $\pm\,1\sigma$ & $95\%$ lower & $95\%$ upper \\
\hline
$\Omega_{\Lambda}$  
& $0.6848$ 
& $0.6893^{+0.0060}_{-0.0058}$ 
& $0.6779$ 
& $0.7009$ \\

$YHe$        
& $0.2478$ 
& $0.2479^{+6.0\times10^{-5}}_{-5.9\times10^{-5}}$ 
& $0.2477$ 
& $0.2480$ \\

$H_0\,[{\rm km\,s^{-1}\,Mpc^{-1}}]$ 
& $67.33$ 
& $67.71^{+0.44}_{-0.44}$ 
& $66.85$ 
& $68.57$ \\

$\sigma_8$          
& $0.8124$ 
& $0.8109^{+0.0059}_{-0.0066}$ 
& $0.7982$ 
& $0.8238$ \\

$\Omega_{\rm rkg}$  
& $0.2645$ 
& $0.2603^{+0.0053}_{-0.0055}$ 
& $0.2496$ 
& $0.2708$ \\

$\Omega_b$          
& $0.04925$ 
& $0.04892^{+0.00050}_{-0.00049}$ 
& $0.04795$ 
& $0.04989$ \\
\hline\hline
\end{tabular}
\end{table}

For the kinetic parameter $\log \beta$ characterizing the dark matter sector, we report only upper limits, given that the posterior distribution exhibits a plateau at low values and decreases sharply beyond a certain threshold. To quantify these upper limits, we compute the values below which $68\%$ and $95\%$ of the posterior samples lie.
\begin{equation}\label{upper_bounds}
\begin{split}
\log \beta &< -6.6890 \pm 0.0031 \quad (\text{68\% credibility})\, ,\\
\log \beta &< -5.3054 \pm 0.0020 \quad (\text{95\% credibility})\, .
\end{split}
\end{equation}

To account for the statistical uncertainty associated with the finite number of posterior samples, we estimate the error on the upper limits using a bootstrap resampling procedure. Specifically, we generate a large number of resampled datasets ($n_{\rm{boot}} = 1000$) by drawing, with replacement, from the original posterior samples, such that each resampled dataset has the same size as the original. For each resample, we compute the corresponding $68\%$ and $95\%$ upper limits. This procedure produces a distribution of upper limits that reflects the variability due to the finite sampling. The standard deviation of these bootstrap distributions is then taken as the uncertainty associated with the upper limits, providing a quantitative estimate of the statistical error without assuming a particular form for the posterior distribution.

\begin{figure}[h!]
    \centering
    \includegraphics[width=7cm]{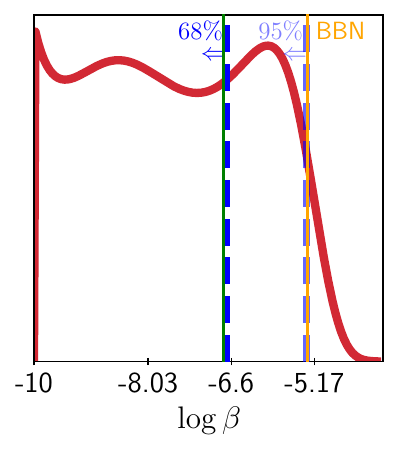}
    \caption{Posterior distribution of the parameter $\log \beta$. Vertical dashed lines indicate the $68\%$ (dark blue) and $95\%$ (light blue) upper limits, while the solid green and orange lines mark the reference value $\beta = 1.806\times 10^{-7}$ and the BBN theoretical constraints respectively. See the text for further details.}  
    \label{1d_beta}
\end{figure}

Figure~\ref{1d_beta} presents the posterior distribution of the kinetic parameter $\log \beta$ separately, highlighting its characteristic plateau and upper-limit behavior. As we mentioned before, the distribution exhibits a plateau at low values followed by a sharp decrease, indicating that smaller values of $\log \beta$ are largely unconstrained by the data, while larger values are strongly limited. To highlight these constraints, we mark the upper limits at $68\%$ and $95\%$ credibility with vertical dashed lines (dark and light blue, respectively). The solid green line indicates the reference value $\beta = 1.806\times 10^{-7}$, corresponding to $\log \beta = -6.7433$. This value lies slightly below the $68\%$ upper bound inferred from the Planck 2018 data, and therefore is consistent with our constraints. Consequently, the upper limit on the present-day root-mean-square velocity of dark matter, $v_\mathrm{rms} \leq 54~\mathrm{m/s}$, reported in~\cite{Armendariz-Picon:2013jej} is compatible with our analysis at the $68\%$ credibility level.

We find that the constraints on the RKG parameter $\beta$ inferred from the Planck 2018 likelihood are fully consistent with the theoretical bound derived from BBN (see Eq.~\eqref{logbeta_bbn}). In particular, the $95\%$ credibility upper limit obtained from cosmological data, $\log \beta < -5.3054 \pm 0.0020$, is in remarkable agreement with the BBN requirement $\log_{10}\beta \lesssim -5.3$, derived from the observational bound on the effective number of relativistic species (see Sec.~\ref{sec:Neff_constraint}). This agreement indicates that the Planck likelihood is already sensitive to the relativistic contribution of the RKG component at early times and effectively enforces the transition of the dark matter fluid into the non--relativistic regime well before recombination. The stronger $68\%$ bound, $\log \beta < -6.6890 \pm 0.0031$, further suggests that current CMB data mildly prefer scenarios in which the relativistic kinetic gas behaves as standard cold dark matter throughout most of the cosmological evolution. Overall, the consistency between the BBN theoretical constraint and the cosmological posterior limits supports the physical viability of the RKG model and confirms that it does not introduce an excessive relativistic energy density during the radiation--dominated era.

\section{Discussion and final remarks}\label{sec:conclusion}

In this work, we have investigated the cosmological viability of a relativistic kinetic gas dark matter scenario by confronting it with current cosmological observations, focusing in particular on CMB measurements from Planck 2018 and complementary large-scale structure data. By embedding the RKG model within a consistent cosmological framework and performing a full Bayesian parameter inference, we have assessed both its impact on the background evolution and its observational signatures at the level of linear perturbations. Our analysis demonstrates that the model smoothly interpolates between a relativistic regime at early times and the standard cold dark matter limit at late times, allowing us to quantify the extent to which present-day data constrain deviations from the $\Lambda$CDM paradigm.

In connection with the analysis of relativistic kinetic gas dark matter presented in~\cite{Armendariz-Picon:2013jej}, our upper bounds~\eqref{upper_bounds} on the kinetic parameter $\log \beta$ can be directly translated into constraints on the present-day dark matter velocity. In particular, we obtain the following upper limits for the characteristic velocity of the relativistic kinetic gas:
\begin{equation}
\begin{split}
|\vec v_0| &\lesssim 61.4 \pm 0.4~\mathrm{m/s}\quad (\text{68\% credibility})\, ,\\
|\vec v_0| &\lesssim 1,484 \pm 6.8~\mathrm{m/s}\quad (\text{95\% credibility})\, ,
\end{split}
\end{equation}
and then, it can be seen that our bounds contain the upper limit of $54$ m/s reported by~\cite{Armendariz-Picon:2013jej}. The significant widening between the $68\%$ and $95\%$ upper limits on the present-day dark matter velocity reflects the highly non-Gaussian nature of the posterior distribution for $\log \beta$. While Planck 2018 data tightly exclude values of $\beta$ large enough to induce substantial relativistic effects in the early universe, they remain fully compatible with the CDM-like regime corresponding to very small kinetic velocities. As a result, moderate departures from the CDM limit are only weakly constrained once the most restrictive region of the posterior is relaxed, leading to a comparatively large difference between the two credibility levels. 
This behavior highlights the fact that linear CMB observables primarily constrain the presence of relativistic corrections at early times, but rapidly lose sensitivity as these effects become subdominant.

Looking ahead, this remaining parameter space can be effectively probed by future cosmological observations sensitive to the non-linear regime of structure formation. In particular, large-scale structure measurements targeting small-scale clustering, redshift-space distortions, and higher-order statistics, as well as weak gravitational lensing surveys such as the Vera C. Rubin Observatory Legacy Survey of Space and Time (LSST)~\cite{LSSTScience:2009jmu,LSSTDarkEnergyScience:2018jkl}, are expected to provide complementary constraints. These probes are directly sensitive to the residual velocity dispersion of dark matter at late times and could therefore significantly tighten the upper limits on $\beta$, enabling a more stringent test of relativistic kinetic gas dark matter scenarios beyond the reach of current CMB data. Importantly, the sensitivity of Planck data to $\log\beta$ highlights the potential of precision cosmology to probe the kinetic nature of dark matter beyond the cold and pressureless approximation. While current constraints are dominated by CMB observations, future large--scale structure surveys will be crucial to further break degeneracies and test the distinctive signatures of relativistic kinetic dark matter on clustering statistics. 

While current constraints are dominated by CMB observations, future large--scale structure surveys will play a key role in breaking degeneracies and probing the distinctive clustering signatures of relativistic kinetic dark matter. This observational progress calls for a more general linear analysis in which perturbations of the phase--space distribution function are explicitly included, enabling a consistent connection between kinetic theory and large--scale structure observables.

From a physical perspective, these results demonstrate that a relativistic kinetic description of dark matter can be consistently embedded within the standard cosmological parameter space. The absence of strong degeneracies with baryonic or dark energy parameters indicates that the RKG model primarily affects the microphysical properties of the dark matter sector, rather than inducing large modifications to the global expansion history.

\appendix
\section{Some comments on the particle mass of the RKG model}\label{app_mass}

At the analytical level, a lower bound on the particle mass cannot be derived from the background evolution alone without specifying the conserved comoving particle number $N$ or, equivalently, a microscopic production mechanism for the dark--matter particles. Nevertheless, the model provides a simple and exact relation linking $m$, $N$, and $\beta$. From the energy density today~\eqref{m_rho}, we have
\begin{equation}
    \rho_{0,\mathrm{DM}} = N\, m\, \sqrt{1+\beta^{2}},
\end{equation}
which implies
\begin{equation}
\label{mass_from_N}
    m = \frac{\rho_{0,\mathrm{DM}}}{N\sqrt{1+\beta^{2}}}.
\end{equation}

Equation~\eqref{mass_from_N} shows explicitly that, for fixed $\beta$, the particle mass is inversely proportional to the conserved comoving number $N$. Therefore, any analytical estimate of a minimum mass necessarily relies on an upper bound for $N$, which must originate from assumptions about the production mechanism of the kinetic gas. From our statistical analysis, we obtain robust upper limits on the kinetic parameter, namely $\log \beta < -6.6890$ at $68\%$ credibility and $\log \beta < -5.3054$ at $95\%$ credibility. These bounds imply that $\beta \ll 1$, such that $\sqrt{1+\beta^{2}} \simeq 1$ to very high accuracy. As a consequence, within the region of parameter space allowed by current data, Eq.~\eqref{mass_from_N} simplifies to
\begin{equation}
    m \simeq \frac{\rho_{0,\mathrm{DM}}}{N}\, ,
    \label{mass}
\end{equation}
showing that cosmological observations effectively constrain the mass only through the specification of the comoving particle number.

For instance, specifying a thermal freeze--out, freeze--in, or any other microscopic production scenario fixes the value of $N$ through the relic abundance calculation, which can then be directly inserted into Eq.~\eqref{mass_from_N} to obtain a numerical estimate for the particle mass. In this sense, the cosmological upper bounds on $\beta$ ensure consistency with the cold dark matter limit, while the mass scale itself remains determined by the underlying particle physics.

In other words, Eq.~\eqref{mass} highlights that different dark matter paradigms correspond to vastly different values of the comoving number density $N$, and that fixing $N$ uniquely determines the implied particle mass. Small values of $N$ correspond to macroscopic dark matter candidates, while large values of $N$ are associated with ultralight particle or field--like descriptions. This mapping provides a convenient unifying language to compare otherwise disparate dark matter models within the same cosmological framework.

Representative values of $N$ and the corresponding particle masses implied by
Eq.~\eqref{mass} are summarized in Table~\ref{tab:m_from_N_DM_models}, where we adopt $\rho_{0,\mathrm{DM}} = 3.7\times10^{76}\,\mathrm{eV\,Mpc^{-3}}\, ,$ consistent with $\Omega_{0,\mathrm{cdm}}\simeq0.26$ and Planck 2018 cosmological parameters. The quoted ranges are intended as order--of--magnitude estimates based on typical values discussed in the literature.

\begin{table}[h!]
\centering
\caption{
Particle mass implied by a given present--day comoving number density $N$
for different dark matter paradigms, assuming
$\rho_{0,\mathrm{DM}} = 3.7\times10^{76}\,\mathrm{eV\,Mpc^{-3}}$ and
$m = \rho_{0,\mathrm{DM}}/N$.
The quoted values are order--of--magnitude estimates.
}
\label{tab:m_from_N_DM_models}
\begin{tabular}{lccc}
\hline\hline
Dark matter paradigm
& $N\;[\mathrm{Mpc^{-3}}]$
& Implied mass $m$ [eV]
& References \\
\hline
Compact objects (PBHs/MACHOs)
& $10^{10}$
& $10^{66}$
& \cite{Carr:2016drx,Sasaki:2018dmp} \\

WIMP--like particles
& $10^{64}$--$10^{66}$
& $10^{12}$--$10^{10}$
& \cite{Bertone:2004pz,Feng:2010gw} \\

Sterile neutrinos
& $10^{72}$--$10^{73}$
& $10^{4}$--$10^{3}$
& \cite{Abazajian:2001nj,Boyarsky:2018tvu} \\

QCD axions
& $10^{80}$--$10^{82}$
& $10^{-4}$--$10^{-6}$
& \cite{Sikivie:2006ni} \\

Ultralight / fuzzy DM
& $10^{98}$
& $10^{-22}$
& \cite{Hu:2000ke,Marsh:2015xka,Matos:2023usa} \\
\hline\hline
\end{tabular}
\end{table}

From a cosmological perspective, Eq.~\eqref{mass} makes explicit that the parameter $N$ controls the microscopic interpretation of the dark matter
sector, independently of the details of the gravitational dynamics. In the context of the RKG model, the parameter $\beta$ governs the late--time modification of structure growth, while $N$ fixes the effective particle mass scale associated with the dark matter component. This separation of roles allows the RKG framework to interpolate continuously between cold, warm, and ultralight dark matter regimes by varying $N$, while keeping the total dark matter abundance fixed.

In summary, while background cosmology alone cannot fix an absolute mass scale for the kinetic gas dark matter particles, the combination of cosmological upper limits on $\beta$ and assumptions about the conserved comoving number density $N$ provides a consistent and flexible framework for deriving physically meaningful mass bounds. Equation~\eqref{mass_from_N} thus constitutes the central link between cosmological constraints and particle properties in relativistic kinetic gas dark matter scenarios.

\section{Use of Lyman--alpha forest likelihood}\label{app_lya}

In order to constrain the parameter $\beta$ using matter power spectrum (MPS) data, we additionally construct a likelihood based on Lyman--$\alpha$ forest measurements reported in Ref.~\cite{Chabanier:2019eai}. This dataset provides access to small comoving scales and thus allows us to test the RKG model through the growth of small--scale structures.

The Lyman--$\alpha$ forest arises from absorption by neutral hydrogen in the intergalactic medium (IGM) along the lines of sight to distant quasars and constitutes a sensitive probe of matter density fluctuations at intermediate and high redshifts, typically in the range $2 \lesssim z \lesssim 5$~\cite{Rauch:1998xn,McDonald:2006qs}. At these epochs, the IGM traces the underlying matter distribution in a relatively clean manner, as non--linear gravitational evolution and complex baryonic processes remain moderately controlled. As a result, statistical measurements of the transmitted--flux power spectrum can be related to the underlying matter power spectrum on small and intermediate comoving scales, roughly $k \sim 0.1$--$10\,h\,\mathrm{Mpc}^{-1}$~\cite{Croft:2000hs,Viel:2013fqw}, which are largely inaccessible to galaxy surveys.

This sensitivity makes Lyman--$\alpha$ observations particularly powerful in testing non--standard dark matter scenarios that predict a scale--dependent suppression of power at small scales, such as warm dark matter or massive neutrinos \cite{Viel:2006yh,Boyarsky:2008xj}. In the case of the RKG model, a non--vanishing value of $\beta$ induces a small effective pressure in the dark sector, leading to a suppression of the growth of small--scale perturbations while leaving large--scale modes essentially unaffected. Consequently, Lyman--$\alpha$ data provide constraints that are complementary to those derived from the CMB and large--scale structure.

The fundamental observable entering Lyman--$\alpha$ forest analyses is the
one--dimensional power spectrum of transmitted--flux fluctuations measured
along individual quasar sightlines,
\begin{equation}
P_{F}^{1\mathrm{D}}(k_\parallel,z)
\equiv
\left\langle
\left| \tilde{\delta}_F(k_\parallel,z) \right|^2
\right\rangle ,
\end{equation}
where $\delta_F \equiv F/\langle F \rangle - 1$ denotes the fractional
fluctuation of the transmitted flux $F = e^{-\tau}$, with $\tau$ the optical
depth of the intergalactic medium. This quantity is directly measured by the
BOSS and eBOSS collaborations over a range of redshifts and scales.

Under the assumption of statistical isotropy, the one--dimensional flux power
spectrum is related to the corresponding three--dimensional flux power spectrum
through
\begin{equation}
P_{F}^{1\mathrm{D}}(k_\parallel,z) = \frac{1}{2\pi}\int_{k_\parallel}^{\infty}
dk\, k\, P_{F}^{3\mathrm{D}}(k,z)\, ,
\end{equation}
which implies that the reconstruction of three--dimensional information from
Lyman--$\alpha$ data is intrinsically indirect.

The three--dimensional flux power spectrum is linked to the underlying matter
power spectrum via a highly non--linear and scale--dependent mapping,
\begin{equation}
P_{F}^{3\mathrm{D}}(k,z) = \mathcal{R}(k,z)\,P_{m}^{3\mathrm{D}}(k,z)\, ,
\end{equation}
where the response function $\mathcal{R}(k,z)$ encapsulates the complex physics
of the intergalactic medium, including hydrodynamical effects, thermal
broadening, pressure smoothing, redshift--space distortions, and astrophysical
feedback. In the Lyman--$\alpha$ likelihood of Ref.~\cite{Chabanier:2019eai}, this mapping is calibrated using suites of high--resolution hydrodynamical simulations assuming a fiducial cold dark matter cosmology.

The matter power spectrum inferred at the redshifts probed by the Lyman--$\alpha$ forest is then mapped onto an equivalent linear matter power spectrum at redshift $z=0$ according to
\begin{equation}
P_{m}^{\mathrm{lin}}(k,z=0) = \left[\frac{D(0)}{D(z)}\right]^2 P_{m}^{\mathrm{lin}}(k,z)\, ,
\end{equation}
where $D(z)$ denotes the linear growth factor computed within the fiducial
cosmological model adopted in the hydrodynamical simulations. The resulting
quantity is commonly referred to as the ``Lyman--$\alpha$ matter power spectrum
at $z=0$'' and constitutes the effective observable constrained by the
likelihood.

It is therefore important to stress that this reconstructed matter power
spectrum is not a direct, model--independent measurement, but rather the output
of a multi--step inference pipeline,
\begin{equation}
P_{F}^{1\mathrm{D}}
\;\longrightarrow\;
P_{F}^{3\mathrm{D}}
\;\longrightarrow\;
P_{m}^{3\mathrm{D}}(z)
\;\longrightarrow\;
P_{m}^{\mathrm{lin}}(z=0),
\end{equation}
whose calibration relies on the assumption of standard cold dark matter.

Accordingly, when applying the Lyman--$\alpha$ likelihood of Ref.~\cite{Chabanier:2019eai} to extended cosmological scenarios, the resulting
constraints should be interpreted as bounds on departures from the fiducial
$\Lambda$CDM (i.e., $\beta=0$) model used in the reconstruction, rather than as
model--independent determinations of the linear matter power spectrum.
In this work, the inferred limits on the RKG coupling parameter $\beta$
therefore quantify the level of deviation from the standard cold dark matter
scenario that remains consistent with the Lyman--$\alpha$ data.

This limitation is closely related to the construction of current Lyman--$\alpha$ likelihoods, which rely on suites of high--resolution hydrodynamical simulations mapping the underlying three--dimensional matter power spectrum onto the one--dimensional transmitted- flux power spectrum measured in quasar absorption spectra; see for instance~\cite{Chabanier:2024knr}. In practice, these simulations are calibrated assuming either the standard cold dark matter paradigm, thermal warm dark matter scenarios, or extensions including massive neutrinos, for which the impact on the linear and mildly non--linear matter power spectrum is well understood and can be accurately interpolated~\cite{McDonald2005,Viel2013,PalanqueDelabrouille2015,Irsic:2017ixq}. As a result, the associated likelihoods are not explicitly calibrated for non--standard dark matter models such as the RKG scenario considered here.

Nevertheless, the use of Lyman--$\alpha$ forest data as an effective constraint on the RKG model is justified under well--defined conditions. In particular, when the coupling parameter $\beta$ is sufficiently small---as independently indicated by CMB--only constraints from \textit{Planck}~2018---the RKG model closely reproduces the expansion history and clustering properties of cold dark matter at redshifts $z \lesssim 5$. In this regime, deviations from standard CDM remain perturbative and do not qualitatively alter the shape of the linear matter power spectrum on the scales probed by the Lyman--$\alpha$ forest.
\begin{figure}[h!]
    \centering
    \includegraphics[width=10cm]{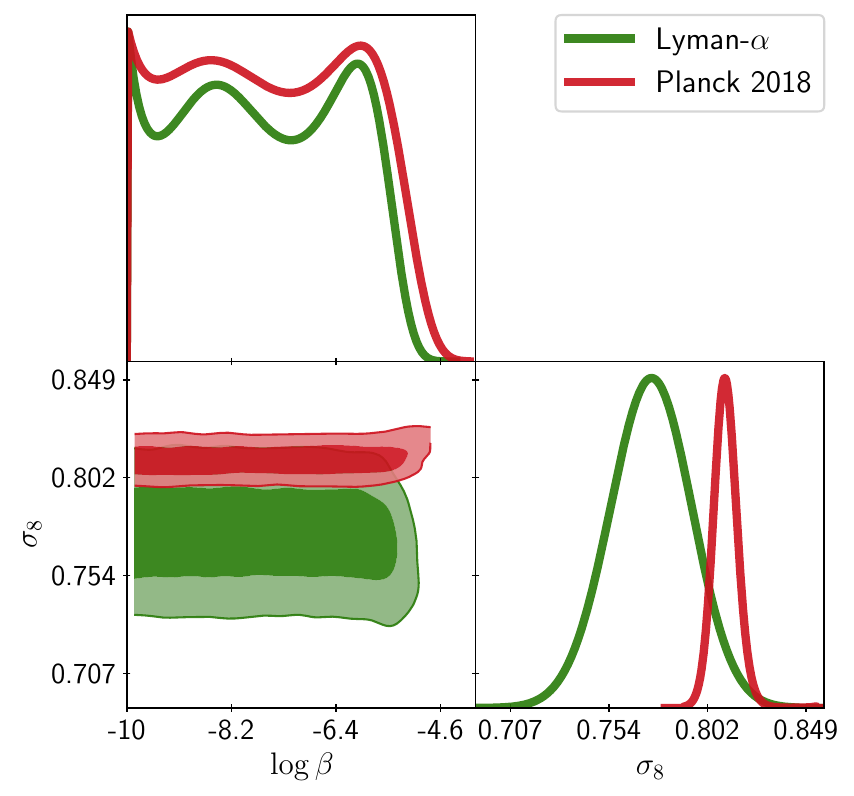}
    \caption{Marginalized constraints in the $\sigma_8$--$\log\beta$ plane for the RKG model from Planck 2018 CMB (green) and Lyman--$\alpha$ data (red), showing a suppression of structure growth and the resulting anticorrelation between $\sigma_8$ and $\beta$}  
    \label{2d_post_lya_cmb}
\end{figure}

Following the same procedure adopted for the CMB constraints, we derive one-- and two--sigma upper bounds from the marginalized posterior distribution. We obtain
\begin{equation}\label{upper_bounds_lya}
\begin{split}
\log \beta &< -6.8828 \pm 0.0027 \quad (68\%~\mathrm{C.L.}),\\
\log \beta &< -5.5596 \pm 0.0016 \quad (95\%~\mathrm{C.L.}).
\end{split}
\end{equation}

A direct comparison between the constraints obtained from CMB (see Eq.~\eqref{upper_bounds}) and those inferred from the Lyman--$\alpha$ flux power spectrum reveals an interesting consistency pattern for the RKG model. The improvement is moderate but systematic: Lyman--$\alpha$ pushes the upper limit on $\beta$ towards smaller amplitudes, in line with the expectation that small--scale structure is more sensitive to the suppression induced by the RKG interaction. In other words, while the CMB is mainly probing the imprint of $\beta$ on early--time growth, the Lyman--$\alpha$ forest traces quasi--linear and mildly non--linear modes where deviations in the matter power spectrum become more prominent. The fact that both probes agree within uncertainties yet peak at slightly different exclusion scales suggests that (i) $\beta$ must be very small to be cosmologically viable and (ii) late--time structure formation provides complementary constraining power beyond primary CMB anisotropies.

Under these assumptions, Lyman--$\alpha$ likelihoods can be interpreted as providing meaningful and conservative constraints on the RKG parameter space, primarily through their sensitivity to the amplitude and scale dependence of matter clustering. We therefore employ Lyman--$\alpha$ data in this work with the explicit caveat that the likelihood is used as an effective constraint, and that a full recalibration based on dedicated hydrodynamical simulations within the RKG framework is beyond the scope of the present analysis.

\bibliographystyle{unsrt}
\bibliography{bib}

\begin{thebibliography}{10}

\bibitem{Planck:2018vyg}
N.~Aghanim et~al.
\newblock {Planck 2018 results. VI. Cosmological parameters}.
\newblock {\em Astron. Astrophys.}, 641:A6, 2020.
\newblock [Erratum: Astron.Astrophys. 652, C4 (2021)].

\bibitem{BOSS:2016wmc}
Shadab Alam et~al.
\newblock {The clustering of galaxies in the completed SDSS-III Baryon
  Oscillation Spectroscopic Survey: cosmological analysis of the DR12 galaxy
  sample}.
\newblock {\em Mon. Not. Roy. Astron. Soc.}, 470(3):2617--2652, 2017.

\bibitem{DES:2021wwk}
T.~M.~C. Abbott et~al.
\newblock {Dark Energy Survey Year 3 results: Cosmological constraints from
  galaxy clustering and weak lensing}.
\newblock {\em Phys. Rev. D}, 105(2):023520, 2022.

\bibitem{Weinberg:2013aya}
David~H. Weinberg, James~S. Bullock, Fabio Governato, Rachel Kuzio~de Naray,
  and Annika H.~G. Peter.
\newblock {Cold dark matter: controversies on small scales}.
\newblock {\em Proc. Nat. Acad. Sci.}, 112:12249--12255, 2015.

\bibitem{KiDS:2020suj}
Marika Asgari et~al.
\newblock {KiDS-1000 Cosmology: Cosmic shear constraints and comparison between
  two point statistics}.
\newblock {\em Astron. Astrophys.}, 645:A104, 2021.

\bibitem{LSSTScience:2009jmu}
Paul~A. Abell et~al.
\newblock {LSST Science Book, Version 2.0}.
\newblock {\em -}, 12 2009.

\bibitem{Euclid:2019clj}
A.~Blanchard et~al.
\newblock {Euclid preparation. VII. Forecast validation for Euclid cosmological
  probes}.
\newblock {\em Astron. Astrophys.}, 642:A191, 2020.

\bibitem{Eifler:2020vvg}
Tim Eifler et~al.
\newblock {Cosmology with the Roman Space Telescope {\textendash} multiprobe
  strategies}.
\newblock {\em Mon. Not. Roy. Astron. Soc.}, 507(2):1746--1761, 2021.

\bibitem{LSSTDarkMatterGroup:2019mwo}
Alex Drlica-Wagner et~al.
\newblock {Probing the Fundamental Nature of Dark Matter with the Large
  Synoptic Survey Telescope}.
\newblock {\em -}, 2 2019.

\bibitem{DES:2024lto}
P.~Shah et~al.
\newblock {The dark energy survey: detection of weak lensing magnification of
  supernovae and constraints on dark matter haloes}.
\newblock {\em Mon. Not. Roy. Astron. Soc.}, 532(1):932--944, 2024.

\bibitem{Bode:2000gq}
Paul Bode, Jeremiah~P. Ostriker, and Neil Turok.
\newblock {Halo formation in warm dark matter models}.
\newblock {\em Astrophys. J.}, 556:93--107, 2001.

\bibitem{Viel:2013fqw}
Matteo Viel, George~D. Becker, James~S. Bolton, and Martin~G. Haehnelt.
\newblock {Warm dark matter as a solution to the small scale crisis: New
  constraints from high redshift Lyman-{\ensuremath{\alpha}} forest data}.
\newblock {\em Phys. Rev. D}, 88:043502, 2013.

\bibitem{Spergel:1999mh}
David~N. Spergel and Paul~J. Steinhardt.
\newblock {Observational evidence for selfinteracting cold dark matter}.
\newblock {\em Phys. Rev. Lett.}, 84:3760--3763, 2000.

\bibitem{Tulin:2017ara}
Sean Tulin and Hai-Bo Yu.
\newblock {Dark Matter Self-interactions and Small Scale Structure}.
\newblock {\em Phys. Rept.}, 730:1--57, 2018.

\bibitem{Boyanovsky:2007ay}
D.~Boyanovsky, H.~J. de~Vega, and N.~Sanchez.
\newblock {Constraints on dark matter particles from theory, galaxy
  observations and N-body simulations}.
\newblock {\em Phys. Rev. D}, 77:043518, 2008.

\bibitem{deVega:2009ku}
H.~J. de~Vega and N.~G. Sanchez.
\newblock {Model independent analysis of dark matter points to a particle mass
  at the keV scale}.
\newblock {\em Mon. Not. Roy. Astron. Soc.}, 404:885, 2010.

\bibitem{Synge:1934zzb}
J.~L. Synge.
\newblock {The energy tensor of a continuous medium}.
\newblock {\em Trans. Roy. Soc. Canada}, III28:127, 1934.

\bibitem{1958PhT....11l..56S}
J.~L. {Synge} and Philip~M. {Morse}.
\newblock {The Relativistic Gas}.
\newblock {\em Physics Today}, 11(12):56, January 1958.

\bibitem{wI63}
W.~Israel.
\newblock Relativistic kinetic theory of a simple gas.
\newblock {\em J. Math. Phys.}, 4:1163--1181, 1963.

\bibitem{CercignaniKremer-Book}
C.~Cercignani and G.M. Kremer.
\newblock {\em The Relativistic Boltzmann Equation: Theory and Applications}.
\newblock {Birkh\"auser}, Basel, 2002.

\bibitem{Vereshchagin-Book}
G.V. Vereshchagin and A.G. Aksenov.
\newblock {\em Relativistic Kinetic Theory with Applications in Astrophysics
  and Cosmology}.
\newblock Cambridge University Press, University Printing House, Cambridge CB2
  8BS, United Kingdom, 2017.

\bibitem{1971grc..conf....1E}
J.~{Ehlers}.
\newblock {General relativity and kinetic theory.}
\newblock In R.~K. {Sachs}, editor, {\em General Relativity and Cosmology},
  pages 1--70, January 1971.

\bibitem{Stewart-Book}
J.M. Stewart.
\newblock {\em Non-Equilibrium Relativistic Kinetic Theory}.
\newblock Lecture Notes in Physics 10, Springer, Berlin, 1971.

\bibitem{Israel:1976tn}
W.~Israel.
\newblock {Nonstationary irreversible thermodynamics: A Causal relativistic
  theory}.
\newblock {\em Annals Phys.}, 100:310--331, 1976.

\bibitem{Sarbach:2013uba}
Olivier Sarbach and Thomas Zannias.
\newblock {The geometry of the tangent bundle and the relativistic kinetic
  theory of gases}.
\newblock {\em Class. Quant. Grav.}, 31:085013, 2014.

\bibitem{rAcGoS2022}
R.~Acu{\~n}a{-}C{\'a}rdenas, C.~Gabarrete, and O.~Sarbach.
\newblock An introduction to the relativistic kinetic theory on curved
  spacetimes.
\newblock {\em General Relativity and Gravitation}, 54(23), February 2022.

\bibitem{Ringstrom-Book}
{H. Ringstr\"om}.
\newblock {\em On the Topology and Future Stability of the Universe}.
\newblock Oxford University Press, Oxford, 2013.

\bibitem{dF16}
D.~Fajman.
\newblock Future asymptotic behavior of three-dimensional spacetimes with
  massive particles.
\newblock {\em Class. Quantum Grav.}, 33(11):11LT01, 2016.

\bibitem{lAdF20}
L.~Andersson and D.~Fajman.
\newblock Nonlinear stability of the {M}ilne model with matter.
\newblock {\em Comm. Math. Phys.}, 378(1):261--298, 2020.

\bibitem{hBdF22}
H.~Barzegar and D.~Fajman.
\newblock Stable cosmologies with collisionless charged matter.
\newblock {\em Journal of Hyperbolic Differential Equations}, 19(04), 2022.

\bibitem{hA11}
H.~Andr\'easson.
\newblock The {E}instein-{V}lasov system/kinetic theory.
\newblock {\em Living Reviews in Relativity}, 14(4), 2011.

\bibitem{Lesgourgues:2011re}
Julien Lesgourgues.
\newblock {The Cosmic Linear Anisotropy Solving System (CLASS) I: Overview}.
\newblock {\em -}, 4 2011.

\bibitem{rMsM87a}
R.~Maartens and S.D. Maharaj.
\newblock Exact inhomogeneous {E}instein-{L}iouville solutions in
  {R}obertson-{W}alker space-times.
\newblock {\em Gen. Rel. Grav.}, 19:499--509, 1987.

\bibitem{rMsM87b}
R.~Maartens and S.D. Maharaj.
\newblock General solution of {L}iouville's equation in {R}obertson-{W}alker
  space-times.
\newblock {\em Gen. Rel. Grav.}, 19:1217--1222, 1987.

\bibitem{rMsM87c}
R.~Maartens and S.D. Maharaj.
\newblock Invariant solutions of {L}iouville's equation in {R}obertson-{W}alker
  space-times.
\newblock {\em Gen. Rel. Grav.}, 19:1223--1234, 1987.

\bibitem{Astorga:2014cka}
Francisco Astorga, Olivier Sarbach, and Thomas Zannias.
\newblock {The evolution of a spatially homogeneous and isotropic universe
  filled with a collisionless gas}.
\newblock {\em J. Phys. Conf. Ser.}, 545(1):012001, 2014.

\bibitem{Astorga:2017yoj}
Francisco Astorga, J.~Felix Salazar, and Thomas Zannias.
\newblock {On the integrability of the geodesic flow on a
  Friedmann--Robertson--Walker spacetime}.
\newblock {\em Phys. Scripta}, 93(8):085205, 2018.

\bibitem{Kuhlen:2012ft}
Michael Kuhlen, Mark Vogelsberger, and Raul Angulo.
\newblock {Numerical Simulations of the Dark Universe: State of the Art and the
  Next Decade}.
\newblock {\em Phys. Dark Univ.}, 1:50--93, 2012.

\bibitem{Vogelsberger:2019ynw}
Mark Vogelsberger, Federico Marinacci, Paul Torrey, and Ewald Puchwein.
\newblock {Cosmological Simulations of Galaxy Formation}.
\newblock {\em Nature Rev. Phys.}, 2(1):42--66, 2020.

\bibitem{Sarbach:2013fya}
Olivier Sarbach and Thomas Zannias.
\newblock {Relativistic Kinetic Theory: An Introduction}.
\newblock {\em AIP Conf. Proc.}, 1548(1):134--155, 2013.

\bibitem{Mangano:2005cc}
Gianpiero Mangano, Gennaro Miele, Sergio Pastor, Teguayco Pinto, Ofelia
  Pisanti, and Pasquale~D. Serpico.
\newblock {Relic neutrino decoupling including flavor oscillations}.
\newblock {\em Nucl. Phys. B}, 729:221--234, 2005.

\bibitem{Lesgourgues:2012uu}
Julien Lesgourgues and Sergio Pastor.
\newblock {Neutrino mass from Cosmology}.
\newblock {\em Adv. High Energy Phys.}, 2012:608515, 2012.

\bibitem{Pitrou:2018cgg}
Cyril Pitrou, Alain Coc, Jean-Philippe Uzan, and Elisabeth Vangioni.
\newblock {Precision big bang nucleosynthesis with improved Helium-4
  predictions}.
\newblock {\em Phys. Rept.}, 754:1--66, 2018.

\bibitem{Moreno:2025wki}
Eladio Moreno and Josue De-Santiago.
\newblock {From Dark Radiation to Dark Energy: Unified Cosmological Evolution
  in K-essence Models}.
\newblock {\em -}, 9 2025.

\bibitem{Ma:1995ey}
Chung-Pei Ma and Edmund Bertschinger.
\newblock {Cosmological perturbation theory in the synchronous and conformal
  Newtonian gauges}.
\newblock {\em Astrophys. J.}, 455:7--25, 1995.

\bibitem{Sachs:1967er}
R.~K. Sachs and A.~M. Wolfe.
\newblock {Perturbations of a cosmological model and angular variations of the
  microwave background}.
\newblock {\em Astrophys. J.}, 147:73--90, 1967.

\bibitem{Chabanier:2019eai}
Sol{\`e}ne Chabanier, Marius Millea, and Nathalie Palanque-Delabrouille.
\newblock {Matter power spectrum: from Ly$\alpha$ forest to CMB scales}.
\newblock {\em Mon. Not. Roy. Astron. Soc.}, 489(2):2247--2253, 2019.

\bibitem{BOSS:2013rlg}
Lauren Anderson et~al.
\newblock {The clustering of galaxies in the SDSS-III Baryon Oscillation
  Spectroscopic Survey: baryon acoustic oscillations in the Data Releases 10
  and 11 Galaxy samples}.
\newblock {\em Mon. Not. Roy. Astron. Soc.}, 441(1):24--62, 2014.

\bibitem{Armendariz-Picon:2013jej}
Cristian Armendariz-Picon and Jayanth~T. Neelakanta.
\newblock {How Cold is Cold Dark Matter?}
\newblock {\em JCAP}, 03:049, 2014.

\bibitem{Planck:2019nip}
N.~Aghanim et~al.
\newblock {Planck 2018 results. V. CMB power spectra and likelihoods}.
\newblock {\em Astron. Astrophys.}, 641:A5, 2020.

\bibitem{Planck:2018lbu}
N.~Aghanim et~al.
\newblock {Planck 2018 results. VIII. Gravitational lensing}.
\newblock {\em Astron. Astrophys.}, 641:A8, 2020.

\bibitem{Beutler:2011hx}
Florian Beutler, Chris Blake, Matthew Colless, D.~Heath Jones, Lister
  Staveley-Smith, Lachlan Campbell, Quentin Parker, Will Saunders, and Fred
  Watson.
\newblock {The 6dF Galaxy Survey: Baryon Acoustic Oscillations and the Local
  Hubble Constant}.
\newblock {\em Mon. Not. Roy. Astron. Soc.}, 416:3017--3032, 2011.

\bibitem{Ross:2014qpa}
Ashley~J. Ross, Lado Samushia, Cullan Howlett, Will~J. Percival, Angela Burden,
  and Marc Manera.
\newblock {The clustering of the SDSS DR7 main Galaxy sample {\textendash} I. A
  4 per cent distance measure at $z = 0.15$}.
\newblock {\em Mon. Not. Roy. Astron. Soc.}, 449(1):835--847, 2015.

\bibitem{Audren:2012wb}
Benjamin Audren, Julien Lesgourgues, Karim Benabed, and Simon Prunet.
\newblock {Conservative Constraints on Early Cosmology: an illustration of the
  Monte Python cosmological parameter inference code}.
\newblock {\em JCAP}, 1302:001, 2013.

\bibitem{Brinckmann:2018cvx}
Thejs Brinckmann and Julien Lesgourgues.
\newblock {MontePython 3: boosted MCMC sampler and other features}.
\newblock {\em -}, 2018.

\bibitem{LSSTDarkEnergyScience:2018jkl}
Rachel Mandelbaum et~al.
\newblock {The LSST Dark Energy Science Collaboration (DESC) Science
  Requirements Document}.
\newblock {\em -}, 9 2018.

\bibitem{Carr:2016drx}
Bernard Carr, Florian Kuhnel, and Marit Sandstad.
\newblock {Primordial Black Holes as Dark Matter}.
\newblock {\em Phys. Rev. D}, 94(8):083504, 2016.

\bibitem{Sasaki:2018dmp}
Misao Sasaki, Teruaki Suyama, Takahiro Tanaka, and Shuichiro Yokoyama.
\newblock {Primordial black holes{\textemdash}perspectives in gravitational
  wave astronomy}.
\newblock {\em Class. Quant. Grav.}, 35(6):063001, 2018.

\bibitem{Bertone:2004pz}
Gianfranco Bertone, Dan Hooper, and Joseph Silk.
\newblock {Particle dark matter: Evidence, candidates and constraints}.
\newblock {\em Phys. Rept.}, 405:279--390, 2005.

\bibitem{Feng:2010gw}
Jonathan~L. Feng.
\newblock {Dark Matter Candidates from Particle Physics and Methods of
  Detection}.
\newblock {\em Ann. Rev. Astron. Astrophys.}, 48:495--545, 2010.

\bibitem{Abazajian:2001nj}
Kevork Abazajian, George~M. Fuller, and Mitesh Patel.
\newblock {Sterile neutrino hot, warm, and cold dark matter}.
\newblock {\em Phys. Rev. D}, 64:023501, 2001.

\bibitem{Boyarsky:2018tvu}
A.~Boyarsky, M.~Drewes, T.~Lasserre, S.~Mertens, and O.~Ruchayskiy.
\newblock {Sterile neutrino Dark Matter}.
\newblock {\em Prog. Part. Nucl. Phys.}, 104:1--45, 2019.

\bibitem{Sikivie:2006ni}
Pierre Sikivie.
\newblock {Axion Cosmology}.
\newblock {\em Lect. Notes Phys.}, 741:19--50, 2008.

\bibitem{Hu:2000ke}
Wayne Hu, Rennan Barkana, and Andrei Gruzinov.
\newblock {Cold and fuzzy dark matter}.
\newblock {\em Phys. Rev. Lett.}, 85:1158--1161, 2000.

\bibitem{Marsh:2015xka}
David J.~E. Marsh.
\newblock {Axion Cosmology}.
\newblock {\em Phys. Rept.}, 643:1--79, 2016.

\bibitem{Matos:2023usa}
Tonatiuh Matos, Luis~A. Ure{\~n}a-L{\'o}pez, and Jae-Weon Lee.
\newblock {Short review of the main achievements of the scalar field, fuzzy,
  ultralight, wave, BEC dark matter model}.
\newblock {\em Front. Astron. Space Sci.}, 11:1347518, 2024.

\bibitem{Rauch:1998xn}
Michael Rauch.
\newblock {The lyman alpha forest in the spectra of quasistellar objects}.
\newblock {\em Ann. Rev. Astron. Astrophys.}, 36:267--316, 1998.

\bibitem{McDonald:2006qs}
Patrick McDonald and Daniel Eisenstein.
\newblock {Dark energy and curvature from a future baryonic acoustic
  oscillation survey using the Lyman-alpha forest}.
\newblock {\em Phys. Rev. D}, 76:063009, 2007.

\bibitem{Croft:2000hs}
Rupert A.~C. Croft, David~H. Weinberg, Mike Bolte, Scott Burles, Lars
  Hernquist, Neal Katz, David Kirkman, and David Tytler.
\newblock {Towards a precise measurement of matter clustering: Lyman alpha
  forest data at redshifts 2-4}.
\newblock {\em Astrophys. J.}, 581:20--52, 2002.

\bibitem{Viel:2006yh}
Matteo Viel, Martin~G. Haehnelt, and Antony Lewis.
\newblock {The Lyman-alpha forest and WMAP year three}.
\newblock {\em Mon. Not. Roy. Astron. Soc.}, 370:L51--L55, 2006.

\bibitem{Boyarsky:2008xj}
Alexey Boyarsky, Julien Lesgourgues, Oleg Ruchayskiy, and Matteo Viel.
\newblock {Lyman-alpha constraints on warm and on warm-plus-cold dark matter
  models}.
\newblock {\em JCAP}, 05:012, 2009.

\bibitem{Chabanier:2024knr}
Sol{\`e}ne Chabanier, Corentin Ravoux, Lucas Latrille, Jean Sexton, {\'E}ric
  Armengaud, Julian Bautista, Tyann Dumerchat, and Zarija Luki{\'c}.
\newblock {The ACCEL2 project: simulating Lyman-{\ensuremath{\alpha}} forest in
  large-volume hydrodynamical simulations}.
\newblock {\em Mon. Not. Roy. Astron. Soc.}, 534(3):2674--2693, 2024.

\bibitem{McDonald2005}
P.~McDonald et~al.
\newblock The linear theory power spectrum from the ly-alpha forest in the
  sloan digital sky survey.
\newblock {\em Astrophys. J.}, 635:761--783, 2005.

\bibitem{Viel2013}
M.~Viel, G.~D. Becker, J.~S. Bolton, and M.~G. Haehnelt.
\newblock Warm dark matter as a solution to the small scale crisis: New
  constraints from high redshift lyman-alpha forest data.
\newblock {\em Phys. Rev. D}, 88(4):043502, 2013.

\bibitem{PalanqueDelabrouille2015}
N.~Palanque-Delabrouille et~al.
\newblock Neutrino masses and cosmology with lyman-alpha forest power spectrum.
\newblock {\em JCAP}, 2015(11):011, 2015.

\bibitem{Irsic:2017ixq}
Vid Ir{\v{s}}i{\v{c}} et~al.
\newblock {New Constraints on the free-streaming of warm dark matter from
  intermediate and small scale Lyman-$\alpha$ forest data}.
\newblock {\em Phys. Rev. D}, 96(2):023522, 2017.

\end{thebibliography}

\end{document}